\documentclass[twoside,11pt]{article}

\usepackage{jmlr2e_clean}

\usepackage{graphics}
\usepackage{graphicx}
\usepackage{caption}
\usepackage{subcaption}
\usepackage{amssymb}
\usepackage{amsmath}
\usepackage{color}
\usepackage{tcolorbox}
\usepackage{mathtools}
\usepackage{placeins}
\usepackage{nicefrac}
\usepackage{booktabs}
\usepackage{mathtools}
\usepackage{dsfont}
\usepackage{multirow}
\usepackage{multirow}
\usepackage{bm, mathrsfs}

\newcommand{\ccF}{{\mathscr F}}
\newcommand{\ccN}{{\mathscr N}}

\usepackage[backgroundcolor=white,bordercolor=orange]{todonotes}

\def\bR{\mathbb{R}}
\def\bN{\mathbb{N}}
\def\cN{\mathcal{N}}

\newcommand{\Ind}{{\mathds 1}}
\newcommand{\ind}[1]{\Ind_{\{#1\}}}

\newcommand{\bx}{\bm{x}}
\newcommand{\ba}{\bm{a}}

\newcommand{\bb}{\boldsymbol{b}}
\newcommand{\VaR}{\operatorname{VaR}}

\DeclareMathOperator{\var}{\mathrm{V}@\mathrm{R}}

\DeclareMathOperator{\LSTM}{LSTM}
\DeclareMathOperator{\NN}{NN}
\DeclareMathOperator{\ER}{ER}

\DeclareMathOperator{\argmin}{argmin}

\jmlrheading{1}{2000}{1-48}{4/00}{10/00}{meila00a}{Ormaniec e.a.}

\usepackage{setspace}

\ShortHeadings{Estimating value at risk: LSTM vs. GARCH}{Ormaniec e.a.}
\firstpageno{1}

\begin{document}

\title{Estimating value at risk: LSTM vs. GARCH}

\author{\name Weronika Ormaniec \email weronika.ormaniec@student.uj.edu.pl\\
       \name Marcin Pitera \email Marcin.Pitera@uj.edu.pl\\
       \addr Institute of Mathematics, Jagiellonian University \\ S. {\L}ojasiewicza 6, 30-348 Krak{\'o}w, Poland
       \AND
       Sajad Safarveisi \email sajad.safarveisi@ionos.com \\
       \addr IONOS, Department of Customer Intelligence \\   Hinterm Hauptbahnhof 5 (Ost), 76137 Karlsruhe, Germany 
\AND
       Thorsten Schmidt \email thorsten.schmidt@stochastik.uni-freiburg.de \\
       \addr Albert-Ludwigs University of Freiburg\\
       Ernst-Zermelo-Str. 1, 79104 Freiburg, Germany.}

\editor{}

\maketitle

\begin{abstract}
Estimating value-at-risk on time series data with possibly heteroscedastic dynamics is a highly challenging task. Typically, we face a small data problem in combination with a high degree of non-linearity, causing difficulties for both classical and machine-learning estimation algorithms. In this paper, we propose a novel value-at-risk estimator using a long short-term memory (LSTM) neural network and compare its performance to benchmark GARCH estimators.
Our results indicate that even for a relatively short time series, the LSTM could be used to refine or monitor risk estimation processes and correctly identify the underlying risk dynamics in a non-parametric fashion. We evaluate the estimator on both simulated and market data with a focus on heteroscedasticity, finding that LSTM exhibits a similar performance to  GARCH estimators on simulated data, whereas on real market data it is more sensitive towards increasing or decreasing volatility and outperforms all existing estimators of value-at-risk in terms of exception rate and mean quantile score. 
\end{abstract}


\section{Introduction}

Risk management is a core topic in both the banking and the insurance industry: risk measures are used for regulatory capital reporting, internal risk monitoring, and other important quantitative-oriented areas such as portfolio optimisation or hedging (see \cite{MFE} or \cite{Car2009} for further details and literature). The proper quantification and estimation of risk is therefore central to these industries. The two most commonly used risk measures are {\it value-at-risk} and {\it expected shortfall}, which have various regulatory frameworks built around them (see e.g. \cite{Bas2016} or \cite{ICS2018}).

Because of its high practical relevance, the estimation of risk and related backtesting procedures are constantly being refined and have received a significant amount of attention. To name only a few key aspects of this important field, we refer to backtesting, elicitability, statistical inference for risk measures, set-valued risk measurement, or risk robustness (see e.g. \cite{ConDegSca2010,Acerbi2014Risk,Ziegel2014,Frank2016,Davis2016,Fissler2016,LauZah2016,KraZah2017,FisHlaRud2021} and references therein).

Due to the dynamic nature of the financial markets, an i.i.d.\ assumption, even on a shorter time period, neglects the stochastic nature of volatility. The seminal work of \cite{Engle1982} and \cite{Bollerslev1987} provides the well-established GARCH framework as a useful modeling class to incorporate this feature in the estimation of risk, e.g. in \cite{duffie1997overview}, \cite{engle2004caviar} and \cite{mcneil2000estimation}.
There are other approaches that deal with the problem of stochastic volatility, and some of them, such as the filtered historical simulation model, are developed specifically for risk measurement (see  \cite{GurMur2015,AngVenBeg2004,SoPhi2006,HarMitPao2006} and references therein). Still, due to various technical constraints linked to the curse of dimensionality, sample efficiency, and dependency modelling problems, the standard historical simulation approach seems to be the most popular choice within the banking sector (see~\cite{EBA2022}). 

Given the tremendous success of deep learning techniques, the quest for an efficient application of machine learning methods in risk estimation has started to become a popular topic recently. The high flexibility of this approach may lead to a viable alternative to classical approaches because it combines the speed and general applicability of standard estimation methods with the time adaptiveness of dynamic models. 
One natural architecture for dynamic models are recurrent neural networks (RNN), which provide a powerful solution to problems that rely on time-dependent data. In the context we consider here, most methods proposed in the literature thus far have been parametric in nature. In other words, the proposed networks typically estimate the parameters of a pre-specified dynamic model (like a GARCH model or a CAViaR model) and parametric plug-in risk estimators are obtained from the network (see \cite{chen2009statistical,wang2019using,wu2019capturing,LiTraWanGerGeo2020} and references therein).

Moreover, for financial time series, neural networks often fail to outperform simple statistical methodologies due to limited learning data (small data context), framework over-complication, or non-standard latent variable processing (see \cite{MakSpiAss2018} for a discussion). The application to risk management also carries the additional problem that the target is to estimate a highly non-linear function of the data which summarises tail information. This additionally reduces the information in the data on the target. 

In this paper, we propose a novel algorithm based on the long short-term memory (LSTM) network introduced in \cite{hochreiter1997long}. The LSTM has already shown good performance for various time series problems including specific GARCH applications (see e.g. \cite{hu2020hybrid} and \cite{kim2018forecasting}). We refer to \cite{greff2016lstm} for an account on the history of LSTM and its various specifications. Other applications may be found in the work of \cite{chalvatzisa2019high}, \cite{wu2019capturing}, and \cite{sirignano2019universal}. 

In particular, we investigate whether a possibly deep LSTM can be efficiently applied to estimating risk on a portfolio level.
To this end, we develop and implement the estimation of value-at-risk for a univariate time series of profits and losses (P\&L) and compare its performance to other estimators. While \cite{chen2009statistical}, \cite{wang2019using}, \cite{wu2019capturing}, \cite{LiTraWanGerGeo2020}, \cite{NguTraGunKoh2022}, and \cite{BarCreGobNguSaa2022} have already shown positive results in this regard, the analysed output was based on a relatively strong assumption imposed on the underlying dynamics, was using external information, or was focused on general dynamics modelling rather than risk quantification. In this paper, we carefully check whether a non-parametric setup can be realised when using LSTM with a scoring function as an objective criterion (see~\cite{Gneiting2011} for more information about point-forecast evaluation). 

In order to overcome the small data problem, we propose a specific \emph{data transformation} procedure which works well even when the underlying data is heteroscedastic. 

Our results indicate that for our purposes a shallow LSTM is fully sufficient. Moreover, the LSTM outperforms standard estimators, such as the empirical estimator or the Gaussian plug-in estimator. The comparison to the GARCH estimators is much more challenging, though. We can show that the  performance of the LSTM is comparable to the performance of the GARCH estimator on simulated data. It even outperforms other estimators when applied to the market data we consider, and, surprisingly, even performs well when the available data is relatively small in machine learning standards, i.e. it contains only around 7\,500 simulated data points, which corresponds to approximately 30 years of daily data. This is remarkable given that the LSTM is a fully non-parametric approach without any structural model information on the underlying data set, which of course is not true for the highly specialised GARCH estimator.

The paper is organised as follows. In Section~\ref{sec:risk} we recall basic concepts. Section~\ref{sec:backtesting} is devoted to backtesting and performance evaluation of risk projections. In Section~\ref{sec:application} we introduce the proposed LSTM estimator. Then, in Section~\ref{sec:empirical}, we assess the performance on both theoretical and simulated data with the focus on a heteroscedastic setting. Finally, we conclude in Section~\ref{S:conclusions}.

\section{The basics of estimating risks}\label{sec:risk}
The goal of this section is to provide a brief introduction to the measurement of risk (via so-called monetary risk measures) and the estimation
of one of the most popular risk measures, value-at-risk. 
Let $(\Omega,\ccF,\mathbb{P})$ be a probability space and let $L^0:=L^0(\Omega,\ccF,\mathbb{P})$ denote the space of all random variables on $(\Omega,\ccF,\mathbb{P})$. The space $L^0$ corresponds to future discounted P\&Ls of financial positions for a given holding period; for simplicity, we assume that the holding period is equal to one day.

To quantify the risk of any random variable from $L^0$, we use a {\it law-invariant monetary risk measure}, i.e. a mapping $\rho\colon L^0\to\bR\cup \{+\infty\}$ which satisfies the following: (1) monotonicity, i.e. $\rho(Z)\geq \rho(Y)$, for $Z,Y\in L^0$ such that $Z\leq Y$; (2) translation invariance, i.e. $\rho(Z+c)=\rho(Z)-c$, for $Z\in L^0$ and $c\in\bR$; (3) normalisation, i.e. $\rho(0)=0$;
(4) law invariance, i.e. $\rho(Z)=\rho(Y)$ if $F_Z\sim F_Y$, i.e. if the distributions of $Z\in L^0$ and $Y\in L^0$ are the same.

Economically speaking, given position $Y$, the value $\rho(Y)$ is used to secure (or insure) potential losses linked to $Y$. Indeed, translation invariance implies that
\begin{equation}\label{eq:cash-add}
  \rho(Y + \rho(Y))=\rho(Y)-\rho(Y)=0,
\end{equation}
thus rendering the position $Y + \rho(Y)$ acceptable or riskless in terms of $\rho$. While most of the  concepts presented in this article are generic and could be applied rather directly to most law-invariant monetary risk measures, we decided to focus on value-at-risk.

For simplicity, let us assume we are given an absolutely continuous random variable $Y\in L^0$, with a strictly positive support, which correspond to the underlying financial position discounted P\&Ls. Since $Y$ is assumed to be continuous, its \emph{value-at-risk} (VaR) at threshold level $\alpha\in (0,1)$ (typical values are $\alpha=1\%$ or $\alpha=5\%$) is simply given by
  \begin{equation}\label{eq:VaR}
    \var_\alpha(Y) := - F^{-1}_Y(\alpha),
  \end{equation}
where $F_Y$ is the distribution of $Y$ under $\mathbb{P}$. Intuitively, the value-at-risk at level $\alpha$ corresponds to the left tail of the distribution at point $\alpha$ (multiplied with $-1$ to achieve a positive capital reserve), i.e.\ losses higher than the value-at-risk occur only with a probability smaller than $\alpha$ (corresponding exactly to $P(-Y>\var_\alpha(Y)) = \alpha$.

Henceforth, let us fix threshold level $\alpha\in (0,1)$ and consider $\rho:=\var_\alpha$ (see~\cite{MFE} for the generic definition of VaR). Note that since $Y\in L^0$ and VaR correspond to a quantile, we always get $\rho(Y)\in \mathbb{R}$, i.e. the risk of $Y$ is finite.

In practical applications, the law of $Y$ (encoded e.g. in $\mathbb{P}$) is unknown and needs to be estimated based on past data. Given $n\in\mathbb{N}$, we use $X:=(X_1,\ldots,X_{n})$ to denote a sample of past $n$ realisations of $Y$. We often refer to $X$ as {\it sample data} or {\it training data}, but please note that at this stage we do not assume that $X$ forms an i.i.d. sample. As usual, to distinguish random variables from deterministic numbers, we denote by small letters specific realisations of random variables. In particular, for $X$ this corresponds to a real vector $x:=(x_1,\ldots,x_{n})$.

A \emph{risk estimator} $\hat\rho\colon \bR^n\to\bR$ is simply a measurable function of the sample data. In particular, we use $\hat\rho(X)$ to denote the estimator of $\rho(Y)$. To reflect the economic interpretation \eqref{eq:cash-add} we sometimes operate directly on 
\begin{equation}\label{eq:Z.biased}
Z:=Y + \hat \rho(X)
\end{equation}
and refer to $Z$ as a \emph{secured position} that is obtained by adding to $Y$ the \emph{estimated} risk capital $\hat \rho(X)$. Note that $Z$ is a random variable and possibly carries risk since $\hat\rho(X)$ is in fact a random variable that estimates the value of $\rho(Y)$.

\begin{remark}[Risk unbiasedness]\label{rem:bias}
From an economic perspective, we would like to find estimator $\hat\rho$ such that the secured position $Z$ defined in \eqref{eq:Z.biased} is non-risky, in the sense that
\begin{equation}\label{eq:unbiased.orig}
\rho(Z)=\rho(Y+\hat\rho(X))=0,
\end{equation}
where $\rho$ measures the true risk of $Z$ under the true (unknown) distribution of $(Y,X)$. The property \eqref{eq:unbiased.orig} could be seen as an empirical counterpart of~\eqref{eq:cash-add} and an equivalent of statistical unbiasedness for risk measures. We refer to~\cite{pitera2018unbiased} for more details.
\end{remark}

\subsection{Estimating value-at-risk in the i.i.d. setting}\label{S:var.iid}
For now let us assume that $X$ is i.i.d. and distributed as $Y$. First, let us introduce the classical non-parametric empirical VaR estimator. Given realised sample $x$, the {\it empirical $\var_\alpha$ estimator} for position $Y$ is given by
\begin{equation}\label{eq:emp estimator}
      \hat{\var}_{\alpha}^{\textrm{emp}} (x):=-x_{(\lfloor n\alpha\rfloor+1)}.
\end{equation}
This is a plug-in estimator in which we plug the estimated (empirical) distribution of $Y$ into \eqref{eq:VaR}. 
Second, we introduce the plug-in Gaussian  estimator. Assume that $Y\sim \cN(\mu,\sigma^2)$, i.e. $Y$ is normally distributed with mean $\mu\in \bR$ and variance $\sigma^2\in\bR_{+}$. Then, for any fixed $\alpha\in (0,1)$, the true risk of $Y$ is equal to $\var_\alpha(Y) = -( \mu + \sigma \Phi^{-1}(\alpha))$.  Clearly, $\mu$ and $\sigma$ are unknown and need to be estimated from the data. Replacing $\mu$ and $\sigma$ with their empirical counterparts
\[
\textstyle \bar{x}:=\frac{1}{n}\sum_{i=1}^{n}x_{i}\quad\textrm{and}\quad \bar{\sigma}(x):=\sqrt{\frac{1}{n-1}\sum_{i=1}^{n}(x_i-\bar{x})^{2}},
\]
we obtain the classical \emph{Gaussian plug-in $\var_\alpha$ estimator} for $Y$ given by
\begin{equation} \label{var: normal}
    \hat\var^{\textrm{norm}}_\alpha(x) := - \Big( \hat\mu(x) + \hat\sigma(x) \Phi^{-1}(\alpha) \Big).
  \end{equation}
  The estimators $\hat{\var}_{\alpha}^{\textrm{emp}}(x)$ and $\hat\var^{\textrm{norm}}_\alpha(x)$ are typically risk biased because they do not satisfy the empirical counterpart of Equation \eqref{eq:cash-add}, i.e. property $\rho(Y + \hat\rho(X))=0$ introduced in Remark~\ref{rem:bias}. Intuitively, this is to be expected: at least, the normal distribution should have been replaced by the $t$-distribution because $\sigma$ is unknown, which is consistent with the prediction interval for the normal distribution (see \cite{Gei1993}). It has been shown in  \cite{pitera2018unbiased} that the \emph{Gaussian unbiased $\var_\alpha$ estimator} given by
\begin{equation}\label{eq:est.var}
\hat{\var}_{\alpha}^{\textrm{u}} (x):=-\left(\bar{x}+\bar{\sigma}(x)\sqrt{\tfrac{n+1}{n}}t^{-1}_{n-1}(\alpha)\right),
\end{equation}
where $t_{n-1}$ stands for the cumulative distribution function of the student-$t$ distribution with $n-1$ degrees of freedom, is unbiased. Of course, the larger the value of $n$, the closer the plug-in estimator will be to the unbiased estimator.

\subsection{Estimating value-at-risk in the presence of heteroscedasticity}\label{S:var.garch}
Financial time series are often non-stationary, e.g. due to volatility changes over time. One popular approach to capture this is to use the class of GARCH models (see \cite{bollerslev1986generalized} or \cite{BauLauRom2006}) amongst many others. Under the usual GARCH setting, the underlying sample $X$ is in fact non-i.i.d., so the classical estimation methodologies presented in Section~\ref{S:var.iid} could no longer be applied directly. In this section we present the typical estimation procedure that is applied in the GARCH setting. 

For simplicity, we use notation $X_{n+1}:=Y$ and assume that $(X_1,X_2,\ldots,X_n,X_{n+1})$ is a subsample of the GARCH($p,q$) process $(X_t)_{t\in\mathbb{Z}}$, i.e. a process that satisfies a recursive dynamics formula given by
\begin{eqnarray}\label{eq-1}
\begin{cases}
X_t &= \sigma_t \varepsilon_t , \\
\sigma_t &= \sqrt{\omega + \sum_{i = 1}^{p}\alpha_i X_{t - i}^2 + \sum_{j = 1}^{q}\beta_j\sigma^2_{t - j}}
\end{cases}, \qquad t\in\mathbb{Z},
\end{eqnarray}
where $(\varepsilon_t)$ is a white noise process, $p,q\in\mathbb{N}$, $\omega\geq 0$, and $\alpha_i,\beta_j\geq 0$, for $(i=1, \ldots, p)$ and $(j = 1, 2,  \ldots, q)$, are some fixed parameters such that $\omega+\sum_{i = 1}^{p}\alpha_i + \sum_{j = 1}^{q}\beta_j < 1$. Note that the parameter restriction ensures that a unique and stationary solution exists, see e.g. \cite{Car2009}. Assuming we know the GARCH($p,q$) parameters and are given information up to time $t=n$, the true conditional VaR for position $Y=X_{n+1}$ is given by
\begin{equation}\label{eq:true.garch}
\var_{\alpha}(Y)=- \sigma F_{\varepsilon}^{-1}(\alpha), 
\end{equation}
where $F_{\varepsilon}$ is the underlying white noise distribution and $\sigma:=\sigma_{n+1}$.

Now, we introduce two GARCH VaR estimators based on different underlying noise assumptions; in both cases we use a pre-fixed $p,q\in\bN$ values.

First, assume that the white noise is Gaussian. In order to estimate \eqref{eq:true.garch} from sample $X$ we need to estimate $\sigma_{n+1}$, which could be seen as a function of parameters $\alpha_i$ and $\beta_j$ as well as the historical data. The typical approach is to use QLE estimation procedure to recover the optimal set of parameters $(\hat\omega, \hat\alpha_i, \hat\beta_j)$ and set initial $\hat\sigma_i$ parameters (for $i=1,2,\ldots,q$)  equal to unconditional long-run variance (see Section II.4.2 in \cite{Car2009} for details). Assuming we are given estimated parameters $\hat\omega$ , $\hat\alpha_i$ and $\hat\beta_j$ that are derived based on sample $x$, we can recover the sequence $(\hat\sigma_i)$ for $i=n,\ldots n-q$ and consequently set
\[
\textstyle \hat \sigma :=  \sqrt{\hat \alpha_0 + \sum_{i = 1}^{p}\hat \alpha_i x_{n+1 - i}^2 + \sum_{j = 1}^{q}\hat \beta_j \hat \sigma^2_{n+1 - j}}.
\]
Then, the {\it Gaussian GARCH VaR estimator} is given by 
\begin{equation}\label{eq:GARCH}
\textstyle  \hat{\var}_{\alpha}^{\textrm{garch-n}}(x) :=  -\hat \sigma \sqrt{\frac{n+1}n} \, t_{n-1}^{-1}(\alpha).
\end{equation}
Note that this is a novel GARCH estimator (introduced in this paper) that differs from the classical Gaussian GARCH plug-in estimator (see Remark~\ref{rem:garch.new} for details).

\begin{remark}[Bias correction for Gaussian GARCH VaR estimator]\label{rem:garch.new}
When defining Gaussian GARCH estimator \eqref{eq:GARCH}, apart from using the plug-in procedure and substituting $\sigma$ with $\hat\sigma$, we also apply the risk bias correction factor in a way similar to that used for the Gaussian distribution, cf. \eqref{var: normal} and \eqref{eq:est.var}. This is done to decrease the size of the estimator's risk bias and to increase its backtesting performance (see Section~\ref{sec:backtesting}). 

We checked this on simulated data: while the estimator \eqref{eq:GARCH} is slightly biased, it outperforms the standard Gaussian GARCH plug-in estimator. 
\end{remark}

Second, let us assume that the underlying white noise follows a t-Student distribution with  unknown parameter $\nu>0$, linked to the degrees of freedom. In this case, we follow a similar procedure as before but take into account another parameter in the QLE estimation procedure, obtaining estimates $(\hat\omega, \hat\alpha_i, \hat\beta_j,\hat\nu)$. Then, the {\it t-Student GARCH VaR estimator} is given by
\begin{equation}\label{eq:GARCH3}
\textstyle  \hat{\var}_{\alpha}^{\textrm{garch-t}}(x) :=  -\hat \sigma \sqrt{\frac{\hat v-2}{\hat v}} \, t_{\hat v}^{-1}(\alpha).
\end{equation}

\section{Backtesting and performance evaluation of risk measures}\label{sec:backtesting}
A well-established means by which to assess the performance of a given estimator is \emph{backtesting}. In a typical backtesting procedure, estimated risks are compared with observed realisations and a performance measure is used to measure the quality (or conservativeness) of the estimation technique. Backtesting in fact very closely resembles the concept of \emph{learning}, whereby \emph{training data} is used to minimise the loss function and \emph{test data} is used to compare the minimiser with realisations in order to select the optimal minimiser. In this paper we focus on the most common approach to backtesting based on a rolling window approach and show how to incorporate it into a machine learning algorithm. 

Hereafter, let us assume that we have data $(X_1,X_2,\dots,X_{m+n})$ of length $m+n\in\bN$ at hand. The value $n\in\mathbb{N}$ corresponds to a (single-point) estimation period length, and $m\in\mathbb{N}$ corresponds to the number of backtests. We use a \emph{rolling window approach:} the $i$th day training dataset (rolling window) is given by
  \begin{equation} \label{def:X}
    W^i := (X_{i},X_{i+1}, \ldots, X_{i+n-1}), \quad i = 1, \ldots, m. 
  \end{equation}
The window $W^i$ is used to compute the $i$-th risk estimator, which we denote by $\hat\rho(W^i)$. This estimator is tested on the following realisation of the portfolio  P\&L,
which is given by
  \begin{equation} \label{def:Y}
  Y^i:=X_{i+n}.
  \end{equation}
The key to most backtesting approaches is to consider  \emph{secured positions} $Z^i$ by adding the estimated risk capital $\hat\rho(W^i)$ to the position: 
 \[
 Z^i:=Y^i+\hat\rho(W^i).
 \]
 The performance is assessed using appropriate performance measures. More specifically, for value-at-risk, this will be the  exception rate test and the mean score based test, which we outline below.

\subsection{Exception rate statistic}\label{S:ER}
For VaR, it is common to count the number of exceptions, also called overshoots, capital breaks or breaches, and confront them with the expected average number of breaches. An exception occurs when the estimated capital is insufficient to cover the realised loss. Namely, in our setting, the \emph{exception rate} statistic is given by
\begin{equation}\label{eq:exception.rate}
\ER(\hat \rho) :=\frac{1}{m}\sum_{i=1}^{m}1_{\{Y^i+\hat\rho(W^i)<0\}}
\end{equation}
(see e.g. \cite{christoffersen2004backtesting}). Ideally, the $\ER$ value should be close to the underlying VaR threshold level $\alpha\in (0,1)$, and \eqref{eq:exception.rate} could be seen as a performance measure that is dual to VaR (see~\cite{MolPit2019}). Note that $\ER(\hat \rho) =\frac{1}{m}\sum_{i=1}^{m}1_{\{Z^i<0\}}$, so that ER could be seen as a function of the secured position. 

More specifically, setting $m=250$ and counting the annual number of capital breaches for regulatory VaR at level $\alpha=1\%$, is the standard regulatory backtest and the core IMA methodology monitoring tool (see \cite{Bas2013,Bas2016}). While in the typical VaR setting the (effective) estimation period must be equal to at least one year ({$n=250$}), there are instances where the number of observations is considerably lower. For instance, this might refer to exotic risks that are quantified using VaR-like measures in the risks not in the VaR (RNIV) environment (see \cite{PRA} for the corresponding local UK regulations or \cite{TRIM} for the EU guidelines linked to risks not in the model engine (RNIME)).

\subsection{Mean quantile score statistic}\label{sec:elicitability}
A recent strand of literature  studies the concept of \emph{elicitability} in the context of point forecast evaluation. Its aim, in our context, is to provide a comparative backtesting method that could be used to confront various competing estimators' performance (see e.g.  \cite{Gneiting2011,Davis2016,Fissler2016}).

The choice of a proper scoring function $S:\bR\times\bR \to \bR_{+}$ is a  key task within the elicitability framework. In the VaR case, the goal of $S$ is to quantify the performance of quantile projections (negative of risk projection) for a given P\&L observation; this value is encoded in arguments of the scoring function. For a proper quantification, one needs to consider function $S$, which is elicitable with respect to the underlying statistic. A typical choice of a quantile consistent scoring function is the \emph{quantile score} given by
\begin{eqnarray}\label{eq:score}
S(x , y) := (\ind{x \geq y} - \alpha)(x - y).
\end{eqnarray}
We refer to \cite{saerens2000building} or \cite{thomson1978eliciting} for more details. As before, the \emph{performance evaluation} is based on comparing the risk predictions $\hat \rho(W^i)$ with P\&L realisations $Y^i$. 
Following  \cite{Gneiting2011}, we consider the \emph{mean quantile score}: 
\begin{equation} \label{eq:average.score}
\bar{S}(\hat \rho) := \frac{1}{m}\sum_{i = 1}^{m}\,S(-\hat \rho(W^i), Y^i).
\end{equation}
Note that for the scoring function introduced in \eqref{eq:score} we have $S(x,y)=S(0,y-x)$, which means that $\bar{S}(\hat \rho)=\frac{1}{m}\sum_{i = 1}^{m}\,S(0, Z^i)$. This allows us to make further economic interpretation of \eqref{eq:average.score}. Namely, we simply check whether the secured position sample $Z^i$ is acceptable, i.e.~it bears no risk. Since the function $S$ could also be used as an objective criterion, we decided to use \eqref{eq:score} as the base building block when applying machine learning methodology. 

Also, we use $\bar{S}$ and $\ER$ to evaluate the efficiency of the estimation methodologies at hand. For completeness, in Example~\ref{ex:2b} we briefly ilustrate the results of a simple backtesting for both exception rate and mean score test statistics.

\begin{example}[Backtesting value-at-risk]\label{ex:2b}
For VaR at level $\alpha=5\%$ we perform the one-day rolling windows procedure with an estimation window equal to $n=50$. The backtest was performed for daily NASDAQ100 returns in the period from 11.03.2005 to 29.01.2021 and for a theoretical simulated sample with the mean and variance fitted to NASDAQ100 data; both time series are of total length $n+m=4\,000$. The risk was estimated using three estimators: the classical  plug-in estimator \eqref{var: normal}, the unbiased estimator \eqref{eq:est.var}, and the empirical estimator \eqref{eq:emp estimator}. The results are presented in Table \ref{t:test1}. 

\begin{table}[tp!]
\centering
{\footnotesize\begin{tabular}{ll*{6}{c}}\toprule
\multirow{2}{*}{Estimator} &&  \multicolumn{3}{c}{NASDAQ} & \multicolumn{3}{c}{Simulated} \\
\cmidrule(lr){3-5}\cmidrule(lr){6-8} 
 &&  exceeds & ER (in \%) & $\bar S$ ($\times 100$) &  exceeds & ER (in \%) & $\bar S$ ($\times 100$)\\ \midrule
Gaussian plug-in & $\hat{\var}_{\alpha}^{\textrm{norm}}$ & 255 & 6.5 & 0.155 & 221 & 5.6& 0.144\\
Empirical & $\hat{\var}_{\alpha}^{\textrm{emp}}$ & 245 & 6.2 & 0.157& 232 & 5.9 & 0.147\\
Gaussian unbiased& $\hat{\var}_{\alpha}^{\textrm{u}}$ & 241  &6.1 & 0.155& 204 & 5.2& 0.144\\ \bottomrule \\
\end{tabular}}
\caption{Backtesting of $\var_{5\%}$ for NASDAQ100 (first column) and for an i.i.d. sample from a normally distributed random variable with mean and variance fitted to the NASDAQ data (second column), both for 4\,000 data points. \emph{Exceeds} reports the number of exceptions in the sample, where the actual loss exceeded the risk estimate. The expected rate closest to theoretical value $5\%$ is reached for the Gaussian unbiased estimator in both cases.}
\label{t:test1}
\end{table}
\end{example}

\section{Estimating risk with LSTM}\label{sec:application}

In this section, we outline how to combine methods presented in Section~\ref{sec:risk} and Section~\ref{sec:backtesting} in order to embed them into a generic machine learning (ML) risk estimation procedure for VaR at level $\alpha\in (0,1)$; the method could easily be modified to account for other risk metrics such as expected shortfall given a proper objective function satisfying the joint elicitability property. The proposed approach is based on a long short-term memory (LSTM) recurrent neural network implementation with {\it mean quantile score} used as an objective criterion. In the proposed estimation framework, we do not rely on the i.i.d. assumption imposed on data and instead allow non-parametric risk estimation. Our goal here is to determine to what extent the neural network is able to recover GARCH-type dynamics on simulated data and whether it outperforms the basic GARCH estimator on market data.

In a nutshell, the class of LSTM networks introduced in  \cite{hochreiter1997long} is an improved subfamily of the recurrent neural networks which is designed to detect dynamic changes by incorporating special layers called {\it LSTM cells}. These cells can be stacked on top of each other and are attached to the input layer of the neural network.
The chain of the LSTM cells may then be followed by an arbitrary number of regular dense layers to produce the output of the network (see \cite{greff2016lstm} for details).

For consistency, we use the notation aligned with that introduced in Section~\ref{sec:backtesting}. Namely, we assume we are given data $(X_1,\ldots,X_{m+n})$ which we split into $m$ batches as in \eqref{def:X}.  For transparency, we split the algorithm description into a few modules (sections), which serve as the building blocks of the estimator.

\subsection{Data transformation}\label{S:input.data}

In this section we describe the data transformation process, which represents a critical step in ensuring that the LSTM algorithm works well on a limited data set. In the case of financial time series, typically only limited data is available. If we consider daily data, a year amounts to approximately 250 data points. Looking at a 10-year history leaves us with only 2500 data points. Typical (relevant) data availability for a financial time series do not exceed 30 years, which corresponds to 7\,500 data points – still, this is clearly a relatively small input when compared to a typical big data implementation, such as that used in image recognition. To overcome this issue, we must enrich the initial input data in such a way that meaningful information is provided directly to the neural network. Namely, to deal with the small sample size, we use \emph{data transformation}: to do this, we add non-linear data transforms based on the first four central moments as well as constant sample mean.

Given a sample $W^i=(X_i,\dots,X_{i+n-1})$, corresponding to the $i$th rolling window data batch, we introduce the \emph{transformed sample} $\tilde W^i=(\tilde W_i,\dots,\tilde W_{i+n-1})$, which is obtained from $W^i$ by adding the first four appropriately scaled moments and the sample mean, i.e.
\begin{eqnarray}\label{eq:tildeX}
\tilde{W}^i = \left(\,\begin{bmatrix}
\bar{W}^i \\ f_1(X_{i}-\bar{W}^i) \\ f_2(X_{i} - \bar{W}^i) \\ f_3(X_{i} - \bar{W}^i)\\ f_4(X_{i} - \bar{W}^i)
\end{bmatrix}\,,\, \begin{bmatrix}
\bar{W}^i  \\ f_1(X_{i + 1}-\bar{W}^i) \\ f_2(X_{i + 1} - \bar{W}^i) \\ f_3(X_{i + 1} - \bar{W}^i) \\ f_4(X_{i + 1} - \bar{W}^i)
\end{bmatrix}\,,\, \ldots\,,\, \begin{bmatrix}
\bar{W}^i \\ f_1(X_{i + n - 1}-\bar{W}^i) \\ f_2(X_{i + n - 1} - \bar{W}^i) \\ f_3(X_{i + n - 1} - \bar{W}^i) \\ f_4(X_{i + n - 1} - \bar{W}^i)
\end{bmatrix}\,\right),
\end{eqnarray}
where $\bar{W}^i = \frac{1}{n}\sum_{j = i}^{i + n - 1}\,X_{j}$ is the mean of the $i$-th window $W^i$, and the functions $f_j$, $j=1,2,3,4$, correspond to the first four centred Chebyshev polynomials, i.e. $f_1(x):=x$, $f_2(x):=2x^2-1$, $f_3(x):=4x^3-3x$, and $f_4(x):=8x^4-8x^2+1$. Note that in \eqref{eq:tildeX}, instead of considering the (central) moment transformations directly, we used the Chebyshev orthogonal basis. This is introduced to improve the stability as well as the learning rate. For example, the core information contained in $(X_i-\hat{x})^2$ and $(X_i-\hat{x})^4$ might be similar, meaning that the algorithm might offset corresponding parameters which would produce a non-robust fit. The introduction of orthogonal polynomials should mitigate such problems.

\begin{remark}
Since $f_1(x)=x$ and the transformed data set contains a sample mean, we are also indirectly including the raw sample $W^i$ in the transformed data set. 
\end{remark}
 
It should be emphasised that the proper data transformation is a critical step in the estimation process in the risk context. Without completing this step, even the recovery of basic data statistics (like variance) becomes a challenge, while the network becomes unable to achieve satisfactory results even for relatively large data sets.

\subsection{Deep LSTM }\label{sec:LSTM1}

In this section we briefly recall the deep LSTM algorithm, referring to \cite{greff2016lstm} for a more thorough description. To capture serial dependence, the recurrent layers of the LSTM network are structured in a specific way.
For simplicity, let us fix the period $i$ and present algorithm processing for data input $ W^i$.
Let  $H\in\mathbb{N}$ denote the number of recurrent layers and $L\in\mathbb{N}$ the number of the  (higher) non-recurrent layers. In each layer $h \in \{1,\dots,L+H\}$ we have $d_h\in\mathbb{N}$ neurons. For $H=1$, this corresponds to a vanilla LSTM while for $H>1$ this refers to a deep LSTM.

Each recurrent layer  $h\in \{1,2\ldots, H\}$ has a specific structure. It contains four components, which we denote by $k\in \{in,f,o,g\}$. The symbols $in$, $f$, and $o$ correspond to the  {\it input gate}, the {\it forget gate}, and the {\it output gate}, respectively, while $g$ denotes the  {\it cell input layer}.  Each component is represented by a map
$\NN^k_{h}\colon \bR^{d_{h} + d_{h - 1}}\to \bR^{d_{h}}$  given by
 \begin{equation}\label{eq:NN}
 \NN^k_{h}(\bx) = \sigma^k ( \ba^k_h +  \bx A^k_h),
 \end{equation}
where $\sigma^{k}$ is a sigmoid function for $k\in \{in,f,o\}$ or tanh function for $k = g$, while
$A_h^k\in \bR^{(d_{h-1} + d_h) \times d_h}$ and $\boldsymbol{a}^k_h \in \bR^{d_h}$ are network calibration parameters.

Next, the LSTM data output recurrent procedure needs to be introduced. For each recurrent layer $h\in \{0,1,\ldots, H\}$ and observation $j\in \{0,1,\ldots n\}$, we use $\LSTM_j^{h}$ to denote recurrent network (data) specific output; for technical reasons we added layer $h = 0$ and observation marker $j = 0$ to the definition. First, we initialise the recursion as follows: for  $h\in \{0,1,\ldots,H\}$ and  $j\in \{0,1,\ldots,n\}$ we set $\LSTM_{0}^h := \mathbf{0} \in \bR^{d_h}$. We set initial {\it cell state} to $c_{0}^h := \mathbf{0} \in \bR^{d_h}$ and plug-in the $i$th (enhanced) data into the network by setting
\[
\LSTM_j^0 := \tilde W_{i+j-1} \in \bR^{d_0}, d_0 > 0.
\]
Second,  the network recursion is specified as follows: given $\LSTM_{j - 1}^h$, $h\in\{1,2,\ldots,H\}$, we set
\[
\LSTM_j^h := o_j^h \odot \sigma(c_j^h),\quad \textrm{for } h=1,2,\ldots,H.
\]
where $\sigma$ is the {\it activation function} tanh, and $a \odot b = (a_1b_1,a_2b_2,\dots)^\top$ is the Hadamard product. Moreover, the $h$th layer gates, cell input, as well as cell state are given as
\begin{align*}
o_j^h &:= \NN_h^o([\LSTM_{j - 1}^h, \LSTM_j^{h - 1}]), \\
f_j^h &:= \NN_h^f([\LSTM_{j - 1}^h, \LSTM_j^{h - 1}]), \\
in_j^h &:= \NN_h^{in}([\LSTM_{j - 1}^h, \LSTM_j^{h - 1}]), \\
g_j^h &:= \NN_h^g([\LSTM_{j - 1}^h, \LSTM_j^{h - 1}]),\\
c_j^h & := f_j^h  \odot c_{j- 1}^h + i_j^h \odot g_j^h,
\end{align*}
where we used $[\cdot, \cdot]$ to denote a partitioned row vector with dimensionality $d_{h} + d_{h - 1}$. Furthermore, for the above equations to be consistent with \eqref{eq:NN}, the matrix $A^k_h$ has to be given in the vectorised form
\[
A^k_h =
\begin{bmatrix}
(A^k_h)_1 \\[3pt]
(A^k_h)_2 \\
\end{bmatrix},
\]
where $(A^k_h)_1$ and $(A^k_h)_2$ are $d_h \times d_h$  and $d_{h - 1} \times d_h$ partition matrices, respectively. Note that for any fixed observation marker $j\in\{1,\ldots,n\}$ one must apply the recursive scheme within the second step to obtain $\LSTM_j^h$ from $\LSTM_j^{h-1}$ for $h=1,2,\ldots,H$.

After stacking the recurrent layers, we provide the output of the LSTM, which gives our risk estimator $\hat \rho$, by passing the output through $L$ non-recurrent layers. The non-recurrent layers are denoted by 
\[
\NN_h(\bx) := \sigma_h (\bb_h +  \bx B_h),
\]
where  $\sigma_h$ is the associated activation function and $\bb_h$ and $B_h$ are vectors and matrices of appropriate size, respectively.
Altogether we arrive at the \emph{deep LSTM risk estimator} defined as
\begin{align}\label{eq:DeepLSTM}
\hat\rho (W^i):=\NN_L \circ \NN_{L-1} \circ \cdots \circ \NN_1(\LSTM^H_{n}),
\end{align}
Of course, the network must be trained for the estimator \eqref{eq:DeepLSTM} to perform well. This corresponds to proper choice of the underlying parameters (e.g. weights) which are fitted based on an objective function value. The fitting procedure is described in detail in the next section.

\subsection{Parameter fitting, objective function, and the final output} \label{S:objective}
A central ingredient in a machine-learning algorithm is the proper objective function specification to allow efficient algorithm training. The deep LSTM risk estimator constructed in
Equation \eqref{eq:DeepLSTM} depends on the network parameters which we  denote by $(A,B)$, where
\[ 
  A :=(\ba_h^k,A_h^k)_{h=1,\dots,H;k\in\{in,f,o,g\}}
\quad\textrm{and}\quad
  B:=(\bb_h,B_h)_{h=1,\dots,L}.
\]
Hence, the LSTM output estimator, i.e. $\hat \rho(W^i)$, is actually a function of $(A,B)$. To emphasise this fact, let us use a modified notation $\hat \rho(W^i| A,B)$ in this section. Given a parameter set $(A,B)$, the LSTM procedure outputs a sequence of projected risk values for all batch samples that are denoted by
\begin{equation}\label{eq:rhoXAB}
\hat \rho(X| A,B):= (\hat\rho(X^1| A,B),\ldots,\hat\rho( X^m| A,B)),
\end{equation}
where $m\in\mathbb{N}$ is the total number of batches used for training. At this point we need to evaluate which parameter choice (A,B) the output \eqref{eq:rhoXAB} is most credible for. To do this, we use {\it mean quantile score} as an objective function. Namely, following the scoring function framework introduced in Section~\ref{sec:elicitability}, and recalling that $Y^i=X_{i+n}$ denotes the next-day P\&L realisation for data batch $W^i$, we want to minimise the following expression:
\begin{align}\label{eq:losswithHequal1}
\bar S(\hat\rho(X|A,B))=&\,\frac{1}{m}\sum_{i = 1}^{m}  \,S(-\hat \rho(W^i|A,B), Y^i)  \notag\\
=&\,\frac{1}{m}\sum_{i = 1}^{m} (\alpha-\ind{Y^i+\hat \rho(\tilde W^i|A,B) \leq 0})\, ( Y^i+\hat \rho(\tilde W^i|A,B)).
\end{align}
The goal of the technical implementation is now to efficiently search for an optimal set of network parameters $(A^*,B^*)$ that produce a sequence of best projected risks, i.e.
\[
(A^*,B^*):=\argmin_{(A,B)} \bar S(\hat\rho(X|A,B)).
\]
Finally, after the optimisation of the objection function \eqref{eq:losswithHequal1} and after determining the optimal network parameters $(A^*,B^*)$, we can define the LSTM VaR estimator by setting
\begin{align}\label{eq:VaRLSTM}
\hat{\var}_{\alpha}^{\textrm{lstm}}(x) := \hat\rho(x|A^*,B^*),
\end{align}
where $\hat\rho(x|A^*,B^*)$ denoted LSTM output for any (single-batch) data set $x=(x_1,\ldots,x_n)$.

\subsection{Technical implementation}\label{sec:implementation}
In order to implement the deep LSTM risk estimator and minimise \eqref{eq:losswithHequal1}, we use \texttt{TensorFlow} (v2.8.0) with the \texttt{Keras} (v2.8.0) API.\footnote{%
The initial parameters of the network were initialised by Xavier-uniform initialisation and the {\it Adam} optimisation algorithm was used for training. We reduced the learning rate by a factor of 0.1 when the metric stopped improving. Moreover, to regularise the network, we used  early stopping  and batch normalisation between the last hidden layer and the output layer. Finally, the input data was rescaled to the range of $[0,1]$.%
}

Given a specific dataset, we follow the usual  scheme by splitting the data into {\it training}, {\it validation}, and {\it test} subsets in the usual 80/10/10 split. The training data is used for calibration of the parameters ($A^*$,$B^*$),  the validation data is used for hyper-parameter tuning and model selection,
and the test data is used for a final out-of-sample performance assessment. We follow the usual  data split. Note that the split is performed without data reshuffling because of the  sequential nature of the data. Finally, to handle potential over-fitting on the validation set, we propose to perform two independent calibrations and to choose the one that performs more consistently on the training and validation subsets. 

We tested several specifications, but it transpired that plain vanilla LSTM with one recurrent layer, $H=1$, is already sufficient for our estimation. Given that the applications we have in mind rely on small data sets, this does not come as a surprise. We also established that five neurons are sufficient to produce acceptable results. For the non-recurrent layers, we achieved good results with one layer having 16 neurons with ReLU activation function and a single value output layer ($L=2$). The proposed architecture is the optimal architecture, clearly outperforming the various other specifications we tested. In particular, we tried adding one more LSTM recurrent layer, but it did not improve the model on test data. Moreover, adding more neurons to the last hidden layer or increasing the number of hidden layers did not improve the results meaningfully.

We also tried using $l_1$ and $l_2$ regularisation on both LSTM and dense layers (see chapter 7 of  
\cite{goodfellow2016deep} for details). However, we discovered that it did not improve the model's performance. Furthermore, we also experimented with other data transformations. We tested using a varied number of Chebyshev's polynomials and passing outputs from other estimators as inputs to our model, i.e. empirical estimation on half or all of the sample. The data transformation presented in Section~\ref{S:input.data} resulted in the most stable model.

Finally, we also tested some modifications of the objective function. Several expressions based on composing the {\it mean quantile score}  with  strictly increasing functions were tested.  This kind of transformation preserves the mean quantile score's economic properties (see \cite{Gneiting2011}), and could be used to additionally penalise $\var$ estimations that stray too far from the realised P\&L value. We also tested the use of additional functions to penalise big deviations from the realised P\&L value. However, we found that an objective function directly based on the mean quantile score as described in Section~\ref{S:objective} performed  best.

\section{Application to simulated and real data}\label{sec:empirical}

The goal of this section is to analyse the performance of the risk estimators on out-of-sample test data using the backtesting performance metrics described in Section~\ref{sec:backtesting}.
To this end, we will consider two reference VaR settings: first, we consider the estimation of VaR at level $\alpha=1\%$ with a learning period equal to $n=250$; second, we consider the estimation of VaR at level $\alpha=5\%$ with a significantly shorter learning period equal to $n=50$.  While the first setting could be considered a standard industry practice for Pillar 1 Market Risk evaluation (see e.g.~\cite{EBA2022}), the second setting is tailored to determine how the estimators perform in a small sample environment. Note that the two settings should have similar statistical power in the sense that one would expect to observe on average 2.5 observations that exceed the theoretical quantile level in both settings (since $250\cdot1\%=50\cdot 5\%=2.5$).

On the one side we use a fully simulated environment and on the other side a real market data set exhibiting heteroscedasticity: in Section~\ref{sec:data:2}, we focus on simulated GARCH and in Section~\ref{sec:data:3} we consider Fama \& French market data. The sample size is $k=\;$7,{}500 corresponding to 30 years of daily quotes. 

For the training of the LSTM estimator, we apply an 80/10/10 split: the training (80\% of the data) and validation sub-sample (10\% of the data) is used for the training of the LSTM estimator while the test sub-sample (10\% of data) is only used for the final out-of-sample performance assessment where we compare it to the other estimation methods. %

The actual backtesting window length is equal to $m=0.1\cdot k-n$ due to the 80/10/10 data split, accounting for approximately 500 (700) backtesting days. To be consistent with the notation introduced in Section~\ref{sec:backtesting}, for $i=1,2,\ldots,m$ we use
\[
W^i=(X_i,\ldots,X_{i+n-1})\quad\textrm{and}\quad Y^i=X_{i+n}
\] 
to denote the $i$th rolling window $W^i$  and the following realised P\&L value $Y^i$.
For $i=1,2,\ldots,m$, we compute the value-at-risk estimators
 \begin{equation*}%
\hat{\var}_{\alpha}^{\textrm{emp}}(W^i), \quad
\hat{\var}_{\alpha}^{\textrm{u}}(W^i),\quad 
\hat{\var}_{\alpha}^{\textrm{garch-n}}(W^i),\quad 
\hat{\var}_{\alpha}^{\textrm{garch-t}}(W^i),\quad \hat{\var}_{\alpha}^{\textrm{lstm}}(W^i),
\end{equation*}
following equations \eqref{eq:emp estimator}, \eqref{eq:est.var}, \eqref{eq:GARCH}, \eqref{eq:GARCH3} and \eqref{eq:VaRLSTM},  respectively; note that for $\hat{\var}_{\alpha}^{\textrm{lstm}}(W^i)$  we use the transformed dataset from~\eqref{eq:tildeX}. Finally, whenever possible, we also include the true  value-at-risk
\[
\var_{\alpha}^{\textrm{true}}(W^i),
\]
which is of course not known in reality but -- as in the case considered here -- is known in simulations and serves as a perfect benchmark.

Once the  estimators are computed, we compare them to the realised P\&L on the next day. Following the framework introduced in Section~\ref{sec:backtesting}, we use two performance metrics that take as their input estimated risks and realised P\&Ls: first, for any estimation method $z \in \{\textrm{emp},\textrm{u},\textrm{garch-n},\textrm{garch-t} ,\textrm{lstm},\textrm{true}\}$ we consider the  average \emph{exception rate}  
   \begin{equation}\label{test:ER}
   \ER(\hat{\var}_{\alpha}^{\textrm{z}}):=\frac{1}{m}\sum_{i=1}^{m}\ind{Y^i+\hat{\var}_{\alpha}^{\textrm{z}}(W^i)<0};
   \end{equation}
   here an \emph{exception} occurs if the estimated risk capital is not sufficient to cover realised losses, i.e.\ when the secure position $Y^i+\hat{\var}_{\alpha}^{\textrm{z}}(W^i)$ suffers a loss (corresponding to the case that the realised loss exceeds the estimated value-at-risk). 
   Since the level of the value-at-risk we target is $\alpha$, we would expect the average exception rate to be close to $\alpha$. Note that average exception rates smaller than $\alpha$ indicate the conservativeness of the estimator; indeed, this is the case when we have fewer exceptions than expected, i.e.\ the estimated value-at-risk is relatively high (and hence conservative).
   
    Second, following  \eqref{eq:average.score} and \eqref{eq:score}, we consider the \emph{mean quantile score} given by
   \begin{equation}\label{test:average score}
   \bar S(\hat{\var}_{\alpha}^{\textrm{z}}) :=\frac{1}{m}\sum_{i = 1}^{m} (\alpha-\ind{Y^i+\hat{\var}_{\alpha}^{\textrm{z}}(W^i) \le 0})\, ( Y^i+\hat{\var}_{\alpha}^{\textrm{z}}(W^i)),
   \end{equation}
   for $z \in \{\textrm{emp},\textrm{u},\textrm{garch-n},\textrm{garch-t},  \textrm{lstm},\textrm{true}\}$. 
   In contrast to the exception rate, for the mean quantile score also the size of the exception matters, thus providng a useful point of comparison for estimators.    
   Note that  smaller values of $\bar S$ indicate the superior performance of the  estimator. 

\subsection{Simulated GARCH  data}\label{sec:data:2}

The simulated data is sampled from eight different GARCH  specifications: a GARCH$(p,q)$ model with $p\in\{1,2,3,4\}$ and  $q = 1$, and either standard normal or $t$-student noise with $\nu=5$ degrees of freedom. The parameters for the GARCH(1,1) model were obtained by fitting the model to the NASDAQ100 index in the period 2016--2021. The other specifications were obtained by manually adding further parameter values while at the same time reducing existing ones. In this way we achieved an increasing level of heteroscedasticity and produced various interesting test examples (see Table \ref{T3:spec} for the exact parameter specification).

\begin{table}[tp!]
\centering
\begin{tabular}{lcccccccc}
\toprule
\multirow{2}{*}{Model} & \multicolumn{6}{c}{GARCH test specification} \\
& $\omega$ & $\alpha_1$ & $\alpha_2$ & $\alpha_3$ & $\alpha_4$ & $\beta_1$ & $\varepsilon_t$ \\
\midrule
GARCH(1,1)-n & 0.000004  & 0.17 & - & - & - &0.8 &$\ccN(0, 1)$ \\
GARCH(2,1)-n & 0.000004 & 0.12 & 0.05 & - & - & 0.8&$\ccN(0, 1)$ \\
GARCH(3,1)-n & 0.000004 & 0.12 & 0.10 & 0.05 & - & 0.7 &$\ccN(0, 1)$ \\
GARCH(4,1)-n & 0.000004  & 0.12 & 0.05 & 0.05 & 0.05 & 0.7&$\ccN(0, 1)$ \\
\midrule
GARCH(1,1)-t & 0.000004  & 0.17 & - & - & - & 0.8&$t_5(0,1)$ \\
GARCH(2,1)-t & 0.000004  & 0.12 & 0.05 & - & - & 0.8&$t_5(0,1)$ \\
GARCH(3,1)-t & 0.000004  & 0.12 & 0.10 & 0.05 & - & 0.7&$t_5(0,1)$ \\
GARCH(4,1)-t & 0.000004  & 0.12 & 0.05 & 0.05 & 0.05 & 0.7&$t_5(0,1)$\\
\bottomrule \\
\end{tabular}

\caption{The chosen  specifications of the simulated GARCH models. Parameters for the GARCH(1,1)-n model were obtained by fitting them to the S\&P500 index in the period 2016--2021 (for $\omega$, $\alpha_1$, and $\alpha_2$, we obtained 95\% parameter confidence intervals  $[-0.02,0.02]$, $[0.00,  0.48]$, and $[ 0.25,  1.34]$). The other specifications were chosen manually.}
\label{T3:spec}
\end{table}

As already stated, we consider $\VaR_{1\%}$ and $n=250$ together with $\VaR_{5\%}$ and $n=50$ on a data set of size 7\,500. The GARCH fitting algorithm (needed to compute e.g. GARCH VaR estimator) is based on the \texttt{arch} (v5.1.0) Python module written by~\cite{bashtage}. 

As already mentioned, we apply an 80/10/10 split to the data set. In the simulated data example we consider here,  to increase statistical efficiency, we resample the last $750$ data points $100$ times and perform the backtesting 100 times. The starting points for the resampling  were obtained using consistent conditional variances from the endpoints of the validation set. Given that, according to Table \ref{T3:spec}, eight parameter specifications were used and we considered $\VaR_{1\%}$ and $\VaR_{5\%}$, altogether $2\cdot 800=\;$1,600 backtesting exercises were performed.

For simplicity, we assume that the GARCH estimators know most of the underlying hyper-parameters. Namely, for Gaussian GARCH(p,1) simulations, we use Gaussian VaR estimator \eqref{eq:GARCH} with pre-fixed parameter $p$, while for t-Student GARCH(p,1) simulations  we use t-Student GARCH VaR estimator \eqref{eq:GARCH3} with pre-fixed parameter $p$. Therefore, for brevity, we refer to the corresponding estimators simply as {\it GARCH estimators} and use $\hat{\var}_{\alpha}^{\textrm{garch}}$ instead of  $\hat{\var}_{\alpha}^{\textrm{garch-n}}$ and  $\hat{\var}_{\alpha}^{\textrm{garch-t}}$. In contrast to the GARCH estimator, the LSTM estimator does \emph{not} know the structure of the underlying noise, nor the $p$ parameter, which complicates the matter significantly.

\subsubsection{Performance assessment} 
The mean and standard deviation of the exception rate $\textrm{ER}$ and the mean quantile score $\bar S$ %
for all estimators (true value-at-risk, empirical, unbiased, GARCH and LSTM)  are provided in Table~\ref{T:garch_0.01} and Table~\ref{T:garch_0.05}. %

\begin{table}[tp!]
\centering
\begin{tabular}{lrrrrr|r}
\toprule
\multirow{2}{*}{Model} & \multicolumn{6}{c}{$\ER$ (in \%)} \\
&  true &  emp &  unbiased &  garch &  lstm &  $B_{\textrm{ER}}$ \\
\midrule
GARCH(1,1)-n & 0.96 {\footnotesize (0.40)} &1.52 {\footnotesize (0.66)} &1.67 {\footnotesize (0.81)} &1.11 {\footnotesize (0.42)} &0.87 {\footnotesize (0.42)} &46\% \\
GARCH(2,1)-n & 1.00 {\footnotesize (0.51)} &1.57 {\footnotesize (0.69)} &1.76 {\footnotesize (0.89)} &1.13 {\footnotesize (0.43)} &1.16 {\footnotesize (0.55)} &54\% \\
GARCH(3,1)-n & 0.94 {\footnotesize (0.49)} &1.57 {\footnotesize (0.64)} &2.08 {\footnotesize (1.08)} &1.12 {\footnotesize (0.49)} &0.55 {\footnotesize (0.42)} &36\% \\
GARCH(4,1)-n & 1.05 {\footnotesize (0.44)} &1.64 {\footnotesize (0.82)} &1.99 {\footnotesize (0.96)} &1.27 {\footnotesize (0.50)} &1.75 {\footnotesize (1.25)} &37\% \\
\midrule
GARCH(1,1)-t & 1.11 {\footnotesize (0.53)} &1.45 {\footnotesize (0.56)} &2.10 {\footnotesize (0.75)} &1.30 {\footnotesize (0.55)} &1.02 {\footnotesize (0.58)} &46\% \\

GARCH(2,1)-t & 0.97 {\footnotesize (0.44)} &1.34 {\footnotesize (0.53)} &1.95 {\footnotesize (0.77)} &1.23 {\footnotesize (0.48)} &1.20 {\footnotesize (0.57)} &44\% \\
GARCH(3,1)-t & 1.03 {\footnotesize (0.39)} &1.40 {\footnotesize (0.61)} &2.02 {\footnotesize (0.93)} &1.32 {\footnotesize (0.41)} &1.36 {\footnotesize (0.81)} &39\% \\
GARCH(4,1)-t & 0.97 {\footnotesize (0.44)} &1.49 {\footnotesize (0.65)} &2.14 {\footnotesize (0.94)} &1.38 {\footnotesize (0.45)} &0.91 {\footnotesize (0.58)} &41\% \\
\bottomrule
\end{tabular}
\\
\vspace{0.2cm}
\begin{tabular}{lrrrrr|r}
\toprule
\multirow{2}{*}{Model} & \multicolumn{6}{c}{$\bar S$ ($\times $10,000)} \\
&  true &  emp &  unbiased &  garch &  lstm &  $B_S$ \\
\midrule
GARCH(1,1)-n & 2.69 {\footnotesize (0.47)} &3.60 {\footnotesize (1.18)} &3.59 {\footnotesize (1.38)} &2.74 {\footnotesize (0.48)} &2.79 {\footnotesize (0.56)} &43\% \\
GARCH(2,1)-n & 2.84 {\footnotesize (0.66)} &4.14 {\footnotesize (1.88)} &4.16 {\footnotesize (2.15)} &2.89 {\footnotesize (0.64)} &3.07 {\footnotesize (0.99)} &30\% \\
GARCH(3,1)-n & 2.52 {\footnotesize (0.71)} &4.11 {\footnotesize (2.21)} &4.31 {\footnotesize (2.71)} &2.57 {\footnotesize (0.71)} &2.82 {\footnotesize (0.94)} &15\% \\
GARCH(4,1)-n & 2.49 {\footnotesize (0.54)} &3.88 {\footnotesize (1.69)} &3.97 {\footnotesize (2.15)} &2.57 {\footnotesize (0.57)} &3.57 {\footnotesize (3.77)} &30\% \\
\midrule
GARCH(1,1)-t & 3.43 {\footnotesize (0.95)} &4.56 {\footnotesize (1.88)} &4.66 {\footnotesize (2.07)} &3.56 {\footnotesize (0.95)} &3.75 {\footnotesize (1.19)} &36\% \\
GARCH(2,1)-t & 3.08 {\footnotesize (0.62)} &4.00 {\footnotesize (1.29)} &4.06 {\footnotesize (1.49)} &3.18 {\footnotesize (0.65)} &3.20 {\footnotesize (0.75)} &54\% \\
GARCH(3,1)-t & 2.75 {\footnotesize (0.80)} &3.90 {\footnotesize (1.66)} &3.97 {\footnotesize (1.87)} &2.89 {\footnotesize (0.88)} &3.30 {\footnotesize (1.32)} &18\% \\

GARCH(4,1)-t & 2.72 {\footnotesize (0.76)} &3.93 {\footnotesize (2.16)} &4.03 {\footnotesize (2.37)} &2.81 {\footnotesize (0.81)} &2.93 {\footnotesize (0.91)} &28\% \\
\bottomrule
\end{tabular}

\caption{The  \emph{exception rate} (top) and the \emph{mean quantile score} (bottom) for estimation $\var_{1\%}$ on simulated GARCH data with  estimation window length $n=250$ for various estimators (true value-at-risk, empirical, unbiased, GARCH and LSTM). 
The data is simulated from each GARCH specification and 100 testing samples are generated, resulting in a total of 800 backtests, each performed on 500 observations. 
Average mean scores (standard deviation in brackets) are presented. 
$B_{\textrm{ER}}$ reports the percentage of samples for which LSTM had an exception rate closer to 1\% when compared to the other estimators, while 
$B_S$ stands for the percentage of samples for which LSTM had the lowest mean score. 
            }
\label{T:garch_0.01}
\end{table}

First, regarding the exception rate in Table \ref{T:garch_0.01}, we recall that a value closest to the true $\alpha=1.00\%$ is best. Moreover, values smaller than $\alpha$ indicate the conservativeness of the estimator. Except for the case of GARCH(4,1)-n, the LSTM estimator outperforms all the classical estimators (empirical, unbiased) and in many cases is surprisingly close to the GARCH estimator (which knows the exact model specifications except for the parameters). Recall that we considered the sample size of $n=250$. In comparison to a sample size of $n=50$, the GARCH estimator can rely on a higher degree of statistical information in the data.
This is confirmed in the GARCH-t cases as well as by the second statistic we report, the mean quantile score.

Regarding the simulations reported in Table \ref{T:garch_0.05}, we recall that an exception rate closest to $5\%$ is best. Now the sample size is $n=50$ while we estimate a less extreme quantile, $\var_{5\%}$. This is certainly more difficult for the GARCH estimator, which is confirmed by the simulation examples: in the simulations, the LSTM estimator is able to outperform the GARCH estimators (and the other classical estimators, of course). As in the previous case, the LSTM estimator is also more conservative than the other estimators, which is appealing from the regulatory capital reporting perspective.
In this example, the LSTM estimator outperforms all the other estimators.

As already mentioned, we performed 800 backtests  for $\var_{1\%}$ and 800 backtests for $\var_{5\%}$. To produce summarising statistics, we counted the number of best performances: the LSTM estimator produced the smallest score in 254 (32\%) cases for $\var_{1\%}$ and 527 (66\%) cases for $\var_{5\%}$. %

\begin{table}[tp!]
\centering

\begin{tabular}{lrrrrr|r}
\toprule
\multirow{2}{*}{Model} & \multicolumn{6}{c}{$\ER$ (in \%)} \\
&  true &  emp &  unb &  garch &  lstm  \\
\midrule
GARCH(1,1)-n & 4.84 {\footnotesize (0.72)} &6.52 {\footnotesize (0.64)} &5.59 {\footnotesize (0.68)} &5.61 {\footnotesize (0.75)} &4.90 {\footnotesize (0.88)} &49\% \\
GARCH(2,1)-n & 5.10 {\footnotesize (0.86)} &6.53 {\footnotesize (0.60)} &5.74 {\footnotesize (0.72)} &5.93 {\footnotesize (0.82)} &5.20 {\footnotesize (1.06)} &40\% \\
GARCH(3,1)-n & 4.85 {\footnotesize (0.80)} &6.64 {\footnotesize (0.70)} &5.87 {\footnotesize (0.66)} &5.98 {\footnotesize (0.77)} &3.68 {\footnotesize (0.89)} &24\% \\
GARCH(4,1)-n & 4.98 {\footnotesize (0.87)} &6.56 {\footnotesize (0.73)} &5.76 {\footnotesize (0.77)} &6.26 {\footnotesize (0.86)} &4.80 {\footnotesize (0.88)} &49\% \\
\midrule
GARCH(1,1)-t & 5.11 {\footnotesize (0.90)} &6.27 {\footnotesize (0.60)} &5.33 {\footnotesize (0.63)} &6.39 {\footnotesize (0.85)} &5.41 {\footnotesize (0.99)} &27\% \\
GARCH(2,1)-t & 4.98 {\footnotesize (0.81)} &6.30 {\footnotesize (0.66)} &5.33 {\footnotesize (0.68)} &6.40 {\footnotesize (0.83)} &5.26 {\footnotesize (1.15)} &35\% \\
GARCH(3,1)-t & 4.91 {\footnotesize (0.72)} &6.43 {\footnotesize (0.65)} &5.46 {\footnotesize (0.71)} &6.70 {\footnotesize (0.76)} &4.24 {\footnotesize (1.18)} &31\% \\
GARCH(4,1)-t & 5.08 {\footnotesize (0.76)} &6.37 {\footnotesize (0.65)} &5.53 {\footnotesize (0.68)} &6.92 {\footnotesize (0.72)} &4.69 {\footnotesize (0.92)} &44\% \\
\bottomrule
\end{tabular}
\\
\vspace{0.2cm}

\begin{tabular}{lrrrrr|r}
\toprule
\multirow{2}{*}{Model} & \multicolumn{6}{c}{$\bar S$ ($\times 10000$)} \\
&  true &  emp &  unb &  garch &  lstm &  $B_S$ \\
\midrule
GARCH(1,1)-n & 10.36 {\footnotesize (1.44)} &11.62 {\footnotesize (1.91)} &11.37 {\footnotesize (1.86)} &10.73 {\footnotesize (1.53)} &10.56 {\footnotesize (1.72)} &82\% \\
GARCH(2,1)-n & 10.97 {\footnotesize (1.86)} &12.40 {\footnotesize (2.43)} &12.15 {\footnotesize (2.36)} &11.40 {\footnotesize (1.92)} &11.28 {\footnotesize (2.19)} &72\% \\
GARCH(3,1)-n & 9.78 {\footnotesize (2.26)} &11.65 {\footnotesize (3.25)} &11.44 {\footnotesize (3.20)} &10.23 {\footnotesize (2.43)} &10.18 {\footnotesize (2.37)} &54\% \\
GARCH(4,1)-n & 9.67 {\footnotesize (1.85)} &11.27 {\footnotesize (2.42)} &11.03 {\footnotesize (2.34)} &10.18 {\footnotesize (1.95)} &10.09 {\footnotesize (2.54)} &75\% \\
\midrule
GARCH(1,1)-t & 10.91 {\footnotesize (2.28)} &12.36 {\footnotesize (2.91)} &12.02 {\footnotesize (2.82)} &11.41 {\footnotesize (2.41)} &11.28 {\footnotesize (2.64)} &73\% \\
GARCH(2,1)-t & 10.48 {\footnotesize (1.74)} &11.85 {\footnotesize (2.32)} &11.49 {\footnotesize (2.18)} &11.00 {\footnotesize (1.82)} &10.83 {\footnotesize (1.98)} &64\% \\
GARCH(3,1)-t & 8.95 {\footnotesize (1.85)} &10.52 {\footnotesize (2.55)} &10.20 {\footnotesize (2.40)} &9.46 {\footnotesize (1.93)} &9.64 {\footnotesize (2.07)} &31\% \\
GARCH(4,1)-t & 9.02 {\footnotesize (1.72)} &10.54 {\footnotesize (2.54)} &10.23 {\footnotesize (2.36)} &9.63 {\footnotesize (1.85)} &9.34 {\footnotesize (1.85)} &76\% \\
\bottomrule
\end{tabular}

\caption{The  \emph{exception rate} (top) and the \emph{mean quantile score} (bottom) for estimating $\var_{5\%}$ on simulated GARCH data with estimation window length $n=50$; compare Table \ref{T:garch_0.01}.
}
\label{T:garch_0.05}
\end{table}

To summarise, the LSTM estimator performs surprisingly well, particularly when recalling that the GARCH estimator is specifically designed for estimating in the context of GARCH-simulated data while the LSTM estimator can be considered a non-parametric estimator. The mean quantile scores of both estimators are very close, suggesting that the two estimators have a similar level of performance. Moreover, both estimators clearly outperform empirical and unbiased risk estimators.

\subsubsection{Simulated paths}

\begin{figure}[tp!]
\hspace{-1.8cm}
\includegraphics[width=1.2\textwidth%
]{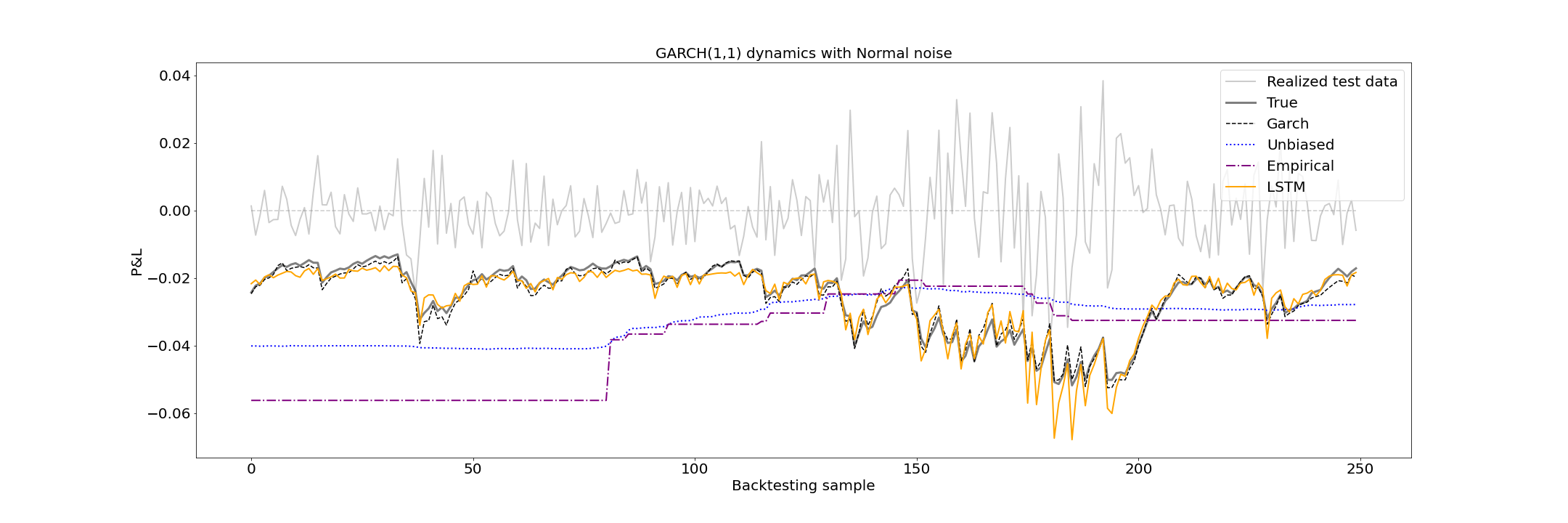} \\
\phantom{.} \hspace{-2.05cm} \includegraphics[width=1.2\textwidth%
]{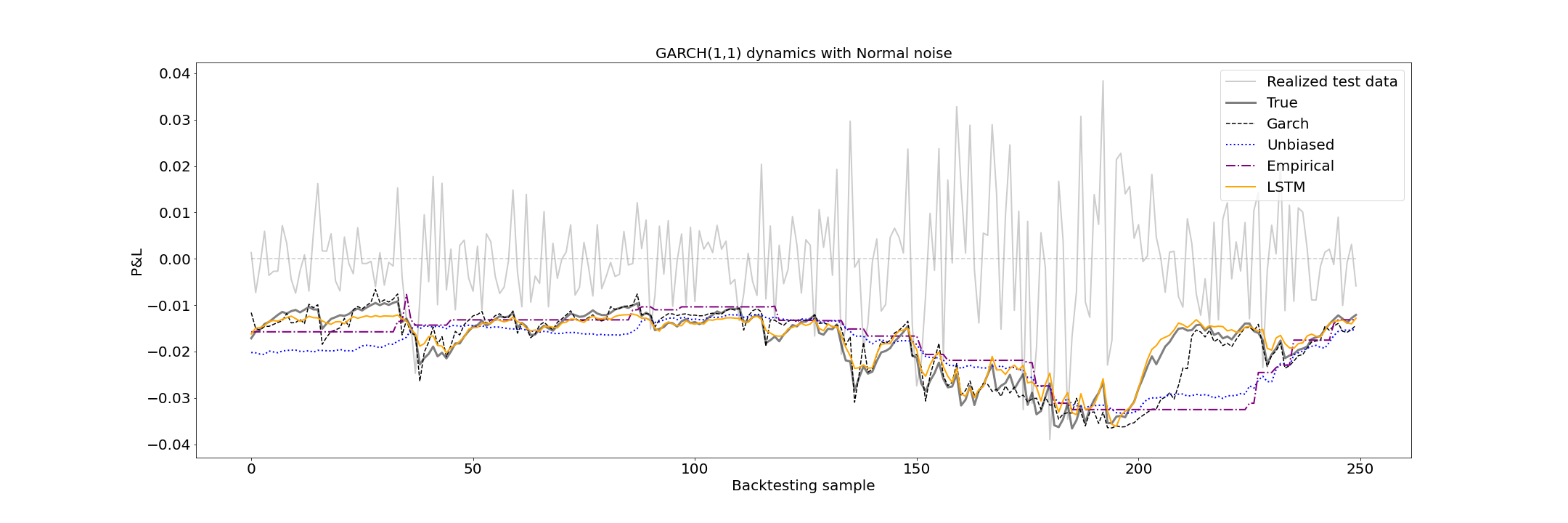}

\caption{250 days of a GARCH(1,1)-n simulation together with risk estimators  for  $\var_{1\%}$  (top) and  for $\var_{5\%}$ (bottom) (multiplied with $-1$). In most cases the LSTM estimator is very close to the GARCH estimator. Top: the LSTM estimator is more sensitive to sudden jumps, i.e. it provides more capital reserve if the P\&L volatility increases. Bottom: %
    the LSTM estimator is closer to true VaR in comparison with the GARCH estimator. %
}
\label{F:garch:1}
\end{figure}

\begin{figure}[t]
\hspace{-1.8cm}
\includegraphics[width=1.2\textwidth]{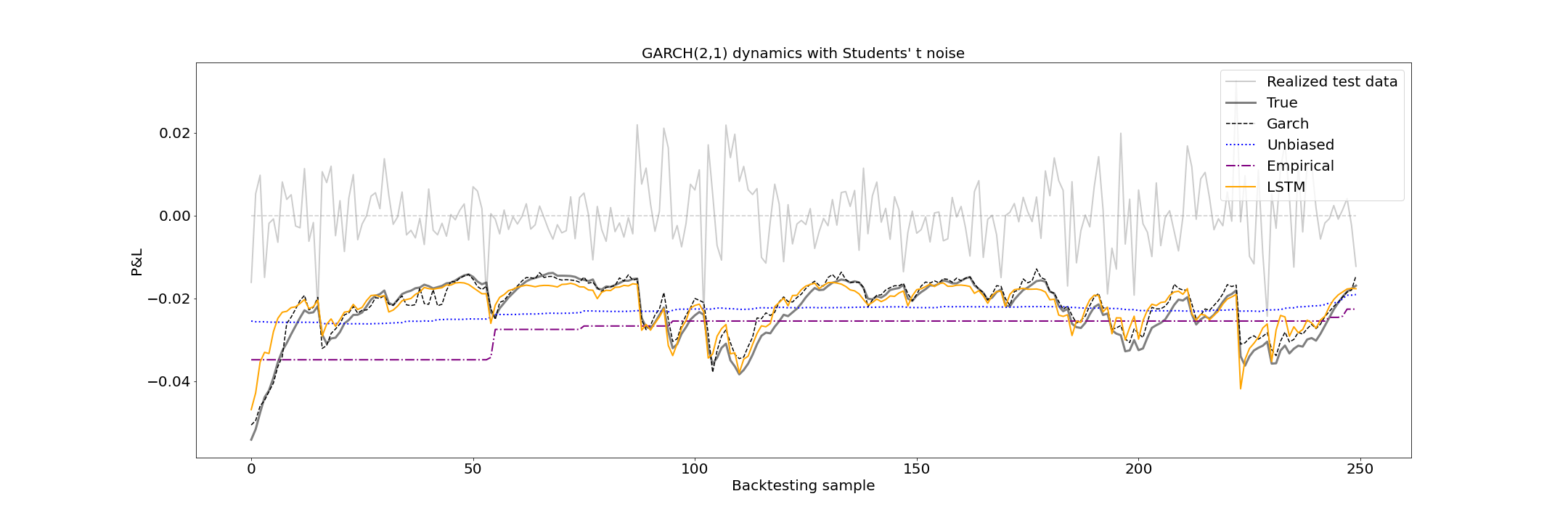}\\
\phantom{.} \hspace{-2.05cm} 
\includegraphics[width=1.2\textwidth]{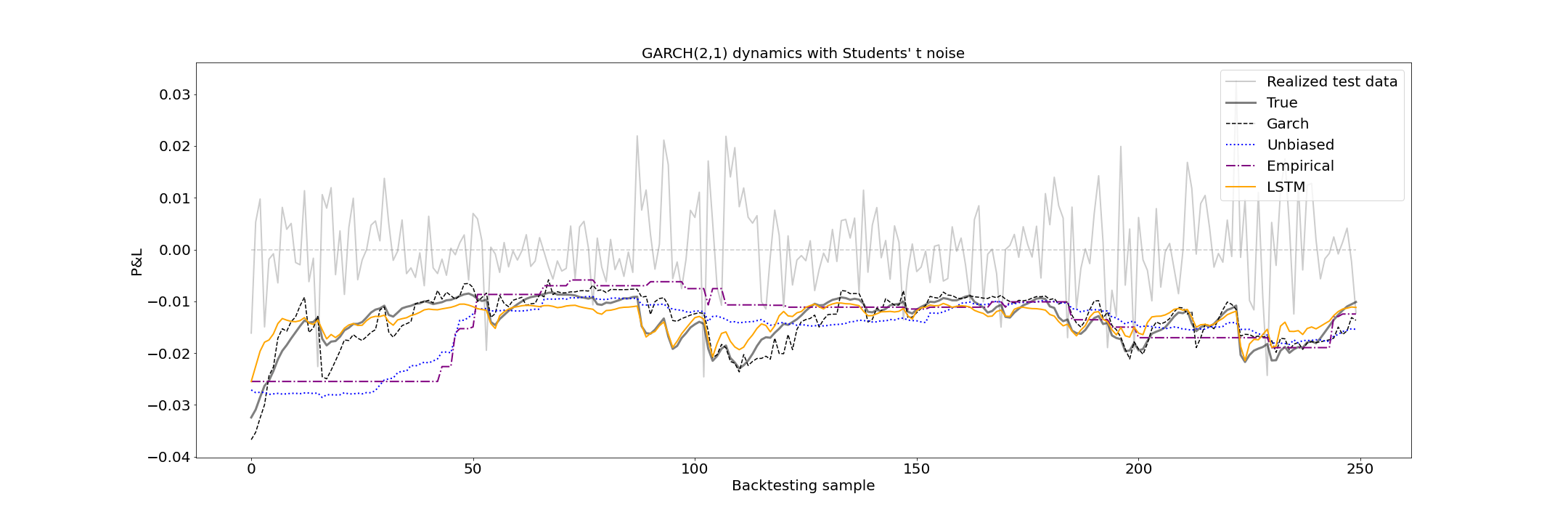}
\caption{250 days of a GARCH(2,1)-t simulation together with risk estimators  for  $\var_{1\%}$  (top) and  for $\var_{5\%}$ (bottom) (multiplied with $-1$). The LSTM estimator is very close to t-Student GARCH estimator, especially in comparison to the  empirical and the unbiased estimator.
}
\label{F:garch:2}
\end{figure}

To gain a better understanding of the behavior of the estimators, we present P\&Ls together with $\var_{1\%}$ and $\var_{5\%}$ estimated values on the last 250 backtesting days in Figure \ref{F:garch:1} and Figure~\ref{F:garch:2}.
These figures provide a clear sense of how the risk estimators behave upon new arising information. Recall for a moment the definition of value-at-risk given in Equation \eqref{eq:VaR}: value-at-risk is $-1$ times the left quantile. For a better comparison to the simulated path, we therefore plot $(-1)$ times the risk estimators, corresponding directly to the left quantile $F^{-1}(\alpha)$.

First, in Figure \ref{F:garch:1} we show estimators for $-\var_{1\%}$ (top) which therefore are significantly lower in comparison to estimators for $-\var_{5\%}$ (bottom) since the higher confidence level of $1-\alpha=99\%$ requires higher capital reserves. It is a striking observation that true value-at-risk together with GARCH and LSTM estimators are all very close to each other, indicating that they are all capable of good performance. 

It can also be observed that the GARCH estimator based on a rolling window of $n=250$ (top) is very capable of capturing underlying dynamics and often even matches the true value-at-risk. The LSTM estimator is, by contrast, slightly more conservative in that it provides a slightly greater amount of capital reserves. This is most easily observed around observation times 170--200 where volatility increases significantly.

 When the rolling window is smaller (bottom, $n=50$), the GARCH estimator experiences greater difficulty in estimating the underlying dynamics while the LSTM estimator is closer to the true value-at-risk, which can be best spotted at observations times 200--220 when the GARCH estimator lags slightly behind and requires some time to adjust back to the true value-at-risk. It is quite surprising that LSTM is able to track this feature so quickly. 
 
 These two plots also show that the classical estimators (empirical and unbiased) struggle to capture the fine-grained volatility movements and follow the true value-at-risk only in a relatively delayed manner. 

In the case of GARCH simulations with $t$-distributed noise, as plotted in Figure \ref{F:garch:2}, we expect an ever better performance from the LSTM estimator since the rougher noise will make it more difficult to estimate the parameters for the GARCH estimator. This was already visible in the statistics presented in Table \ref{T:garch_0.01}.

This is indeed the case: in the first case (top, estimation of $\var_{1\%}$), both estimators are very close. Again, in the case of a spike, the LSTM estimator is slightly more conservative than the GARCH estimator. Meanwhile, in the case with the smaller sample size (bottom), the LSTM estimator seems to be fractionally closer to the true value-at-risk, which confirms our previous findings.

\FloatBarrier

\subsection{Empirical Fama \& French dataset} \label{sec:data:3}
For empirical analysis we take Fama \& French datasets that include returns of 25 portfolios formed on book-to-market and operating profitability on two different periods (see \cite{FamFre2020}).

In light of the pandemic, we consider two separate time periods to determine how the stressed market conditions in the last two years impact the estimator performance. The \emph{first period} ranges from 02.06.1989 to 08.03.2019 while the \emph{second period} ranges from 20.05.1992 to 28.02.2022.

In the first period, the last 10\% of observations correspond to a non-stressed environment, when no sudden regime switch is observed. On the other hand, in the second period a sudden change in dynamics  could be observed for the last 10\% of the data, which is a result of the COVID-19 pandemic outbreak. By considering these two cases, we can evaluate the performance of the estimators in both non-stressed and stressed market conditions. The total number of observations in both datasets, i.e. daily returns rates for each portfolio, is equal to 7\,500, which is consistent with the setting from Section~\ref{sec:data:2}. 

As in the previous section, we consider $\VaR_{1\%}$ with $n=250$ as well as $\VaR_{5\%}$ with $n=50$.
Given that the distribution of the underlying noise is unknown, we include both the Gaussian GARCH  and t-Student GARCH estimators. Since on the S\&P 500 data the estimation yielded a GARCH(1,1) structure, we stick to this specification for the following application. Of course, in this section the results for true risk are unknown.

\begin{figure}[tp!]

\hspace{-1.8cm}\includegraphics[width=1.2\textwidth]{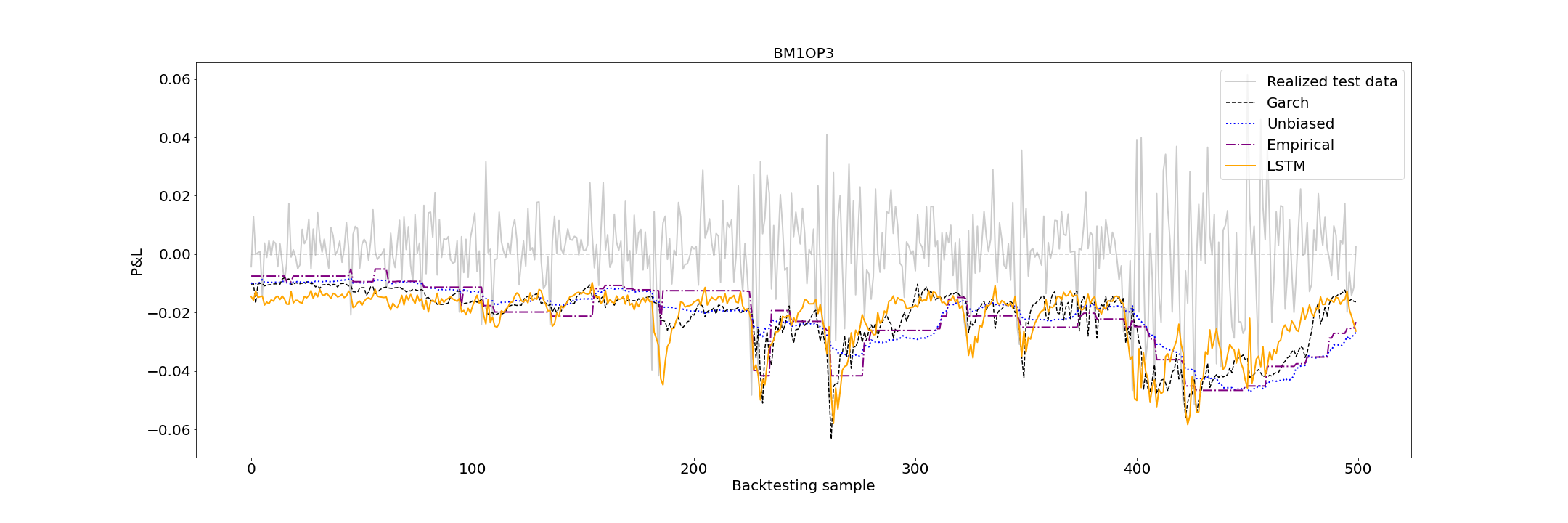} %
\caption{\emph{Non-stressed period (from 14.03.2017 to 08.03.2019):}The plot presents the last 500 realised observations and risk projections of an exemplary portfolio from the non-pandemic Fama \& Fench dataset for $\var_{1\%}$. GARCH in this plot refers to GARCH-t which outperforms GARCH-n. }
\label{F:fama1}
\end{figure}

\begin{table}[htp!]
\phantom{.}\hspace{-0.8cm}
\begin{tabular}{lcccccccccccc}
\toprule
\multirow{2}{*}{Data} & \multicolumn{5}{c}{$\ER$ (in \%) } &  \multicolumn{5}{c}{$\bar S$ ($\times 10000$)} \\
\cmidrule(lr){2-6}\cmidrule(lr){7-11}
&  emp & u & garch-n & garch-t & lstm &  emp & u & garch-n & garch-t & lstm \\
\midrule
BM1OP2   &     2.00 &     5.00 &     3.20 &     2.80 &{\bf 1.00}   &      5.71 &     7.68 &     5.99 &     5.57 &{\bf 5.22}     \\
BM1OP3   &     2.00 &     4.00 &     2.60 &     2.00 &{\bf 1.60}   &      5.24 &     6.56 &     5.41 &     4.93 &{\bf 4.39}     \\
BM1OP4   &{\bf 2.00} &     3.80 &     3.00 &     2.20 &{\bf 2.00}  &      4.13 &     5.43 &     4.66 &     4.14 &{\bf 3.80}     \\
BM2OP1   &{\bf 1.60} &     5.00 &     3.40 &     2.60 &     0.20   &      5.26 &     6.66 &     5.76 &     5.60 &{\bf 4.64}     \\
BM2OP2   &     2.40 &     5.20 &     3.80 &     2.20 &{\bf 0.80}   &      3.92 &     5.31 &     4.89 &     4.28 &{\bf 3.33}     \\
BM2OP3   &{\bf 1.80} &     3.60 &     3.00 &     2.20 &     2.00   &      4.29 &     5.74 &     5.04 &{\bf 4.12} &     4.37     \\
BM2OP4   &{\bf 1.20} &     4.00 &     2.60 &     1.60 &{\bf 0.80}  &      3.34 &     4.08 &     3.20 &     2.94 &{\bf 2.78}     \\
BM2OP5   &{\bf 1.20} &     3.60 &     2.40 &     2.20 &     2.00   &      3.63 &     4.34 &     3.98 &     3.64 &{\bf 3.11}     \\
BM3OP1   &{\bf 0.60} &     2.00 &{\bf 1.40} &{\bf 1.40} &    0.40  &      3.19 &     3.52 &     3.16 &     3.38 &{\bf 2.95}     \\
BM3OP2   &{\bf 1.60} &     3.40 &     2.60 &     2.20 &     1.80   &      3.84 &     4.68 &     3.93 &     3.64 &{\bf 3.46}     \\
BM3OP3   &{\bf 1.40} &     3.60 &    2.20 &{\bf 1.40} &{\bf 1.40}  &      3.36 &     4.20 &     3.64 &     3.50 &{\bf 2.88}     \\
BM3OP4   &     2.00 &     3.00 &     2.60 &     2.20 &{\bf 1.20}   &      3.50 &     3.93 &     3.61 &     3.42 &{\bf 2.85}     \\
BM3OP5   &{\bf 1.60} &     3.00 &     2.60 &     1.80 &     2.20   &      4.84 &     5.51 &     5.15 &{\bf 4.74} &     4.95     \\
BM4OP1   &     1.60 &     3.00 &     1.80 &{\bf 1.20} &{\bf 1.20}  &      4.31 &     4.49 &     4.20 &     3.92 &{\bf 3.83}     \\
BM4OP2   &     1.60 &     3.20 &     2.80 &     1.80 &{\bf 1.20}   &      3.56 &     4.42 &     4.02 &     3.66 &{\bf 3.33}     \\
BM4OP3   &{\bf 1.20} &     2.40 &     2.00 &     1.40 &{\bf 0.80}  &      3.68 &     3.95 &     3.78 &     3.65 &{\bf 3.16}     \\
BM4OP4   &{\bf 1.20} &     3.00 &     3.20 &     2.20 &     1.80   & {\bf 3.84} &     4.70 &     4.65 &     4.31 &     4.03     \\
BM4OP5   &{\bf 1.00} &     2.00 &     2.40 &     2.00 &     0.60   &      4.60 &     4.74 &     4.99 &     4.89 &{\bf 4.39}     \\
BM5OP2   &{\bf 1.00} &     2.40 &     2.00 &     1.80 &     1.60   &      4.36 &     5.15 &     5.22 &     5.08 &{\bf 4.19}     \\
BM5OP3   &{\bf 1.20} &     3.20 &     2.40 &     1.80 &     2.00   & {\bf 4.24} &     4.97 &     5.07 &     4.79 &     4.38     \\
BM5OP4   &     2.00 &     2.40 &     2.60 &     2.40 &{\bf 1.00}   &      10.71 &     12.72 &     10.92 &{\bf 8.86} &     8.89  \\
HiBMHiOP &     1.60 &     1.40 &{\bf 1.20} &     0.60 &     0.40   &      5.33 &{\bf 5.27} &     5.54 &     5.68 &     5.47   \\
HiBMLoOP &     2.40 &     2.60 &     2.40 &     2.20 &{\bf 0.60}   &      3.77 &     4.41 &     3.87 &     3.86 &{\bf 3.34}   \\
LoBMHiOP &     1.80 &     4.20 &     2.40 &     2.20 &{\bf 1.40}   &      3.64 &     4.82 &     4.18 &     3.85 &{\bf 3.30}   \\
LoBMLoOP &     2.00 &     4.40 &     3.60 &     2.40 &{\bf 1.00}   &      5.36 &     6.92 &     6.18 &     5.48 &{\bf 4.67}   \\
\midrule
Bold count  &14 &0 &2 &3 &14  &2 &1 &0 &3 &19 \\
\bottomrule
\end{tabular}
\caption{\emph{Non-stressed period:} The  \emph{exception rate} (top) and the \emph{mean quantile score} (bottom) for estimation $\var_{1\%}$ on the Fama \& French data set with  estimation window length $n=250$ for various estimators (empirical, unbiased, GARCH-n, GARCH-t and LSTM). For the exception rate (ER), values closest to $1\%$ perform best (marked in bold). As summary statistics we provide the bold count. For the mean quantile score, lowest values perform best. }
\label{T:fama1}
\end{table}

\subsubsection{First period (non-stressed)}
The first period is the non-stressed period and we plot the last 500 realised P\&Ls of an exemplary portfolio together with the estimators in Figure \ref{F:fama1}. 
First, heteroscedasticity is clearly visible in the time period. Moreover, the beginning of the period is quieter while there are two clusters of higher volatility, one between 200 and 300 and a further one from 400 onwards. The latter corresponds to reactions to the trade war between China and the United States and poses a significant challenge to the estimators. 

Before studying the estimators in this figure in more detail, we first analyse the results in Table~\ref{T:fama1}. 
In contrast to the results on the simulated GARCH data, we have a surprisingly different picture: first, on the exception rate statistic, the empirical estimator performs much better in comparison to its performance on the simulated data. Second, the GARCH estimators perform much worse on the real data. Third, the GARCH-t estimator performs slightly better than the GARCH-n estimator. Fourth, the LSTM estimator outperforms the other estimators in 14 cases -- just as the empirical estimator does. If it is outperformed by the empirical estimator, it typically exhibits rather conservative behaviour that is evident in lower ER statistics. 

The picture changes, though, if one considers the mean quantile score $\bar S$. Here, the LSTM estimator clearly outperforms all the other estimators (19 out of 25 instances). This shows that the LSTM estimator is very well balanced and provides excellent risk estimation. 

The estimation of $\var_{5\%}$ produces similar results and is therefore not presented. %
Now we turn back to Figure \ref{F:fama1}. Here one can see that the empirical quantile roughly follows the volatility structure, which leads to a surprisingly good performance in terms of the exception rate statistics. However, one can also see that, most notably, in the last cluster with high volatility the empirical estimator follows the development of the volatility with delay. From approximately the 400th observation, it takes a number of days until it increases to a higher level of capital reserve, after which it decays much slower in comparison to the LSTM estimator when volatility decreases again. 

Moreover, it can be seen that the LSTM and GARCH estimators are relatively similar. One spike around 180 is not identified by the GARCH estimator. Moreover, in the volatility cluster around 400, the LSTM estimator picks up the development faster and, after the phase of high clustering, decays faster to a normal level. This corresponds exactly to our previous findings and illustrates how the LSTM estimator outperforms all other estimators regarding the mean quantile score. It might be recalled at this point that the exception rate only counts the number of exceptions or overshoots (when there is insufficient capital), whereas the mean quantile score also takes the size of the overshoot into account and hence provides more precise statistics on the performance of the estimator.

\subsubsection{Second period (stressed)}
\begin{figure}[htp!]
\hspace{-1.8cm}\includegraphics[width=1.2\textwidth]{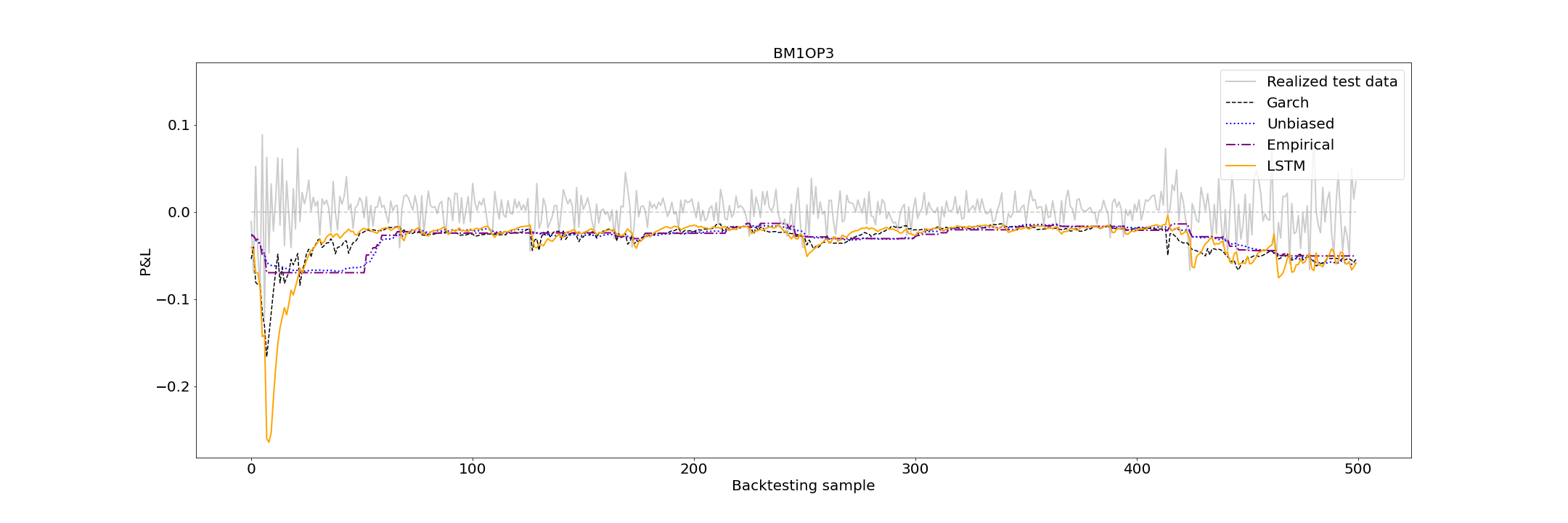}
\caption{\emph{Stressed period (from 06.03.2020 to 28.02.2022):} the plot presents the last 500 realised observations and risk projections for two exemplary portfolios from the pandemic Fama \& Fench dataset for $\var_{1\%}$ (top row) and  $\var_{5\%}$ (bottom row). We decided to not present the results for the $\hat{\var}_{\alpha}^{\textrm{garch-n}}$ estimator because its dynamics are similar to $\hat{\var}_{\alpha}^{\textrm{garch-t}}$ but it has a higher mean score, see Table~\ref{T:fama1}.}
\label{F:fama2}
\end{figure}

The second period is the stressed period from 20.05.1992 to 28.02.2022, which includes the pandemic crisis, which, in March 2020, triggered a free fall in the stock markets. In the last 500 realised P\&Ls of an exemplary portfolio in Figure \ref{F:fama2} this crisis period is just at the left side. In comparison to Figure \ref{F:fama1}, the scale changed from $[-0.06,0.06]$ to $[-1,1]$, and the shown volatility cluster in the stressed period is certainly more severe than the non-stressed period. 

Particularly in the pandemic crisis, the empirical estimator requires a certain amount of time until it rises to higher levels, which then remain comparatively low. The GARCH and LSTM estimators, however, quickly pick up the stressed scenario and -- in a seemingly important development -- cool down much slower. Both are important properties in practice. As we have already noted, in the non-stressed period, the exception rate will not be able to identify this behaviour while the mean quantile score will. 

Not trained on such extreme scenarios, the LSTM estimator overshoots to a relatively high level in the pandemic crisis. While the GARCH estimator picks up the crisis rising only a little slower, it does not require those high levels of capital, which is certainly advantageous in practical applications. 

Visiting the numerical results in Table~\ref{T:fama2}, we immediately see that the stressed period is a challenge to all estimators. Regarding the exception rate, the empirical estimator clearly performs best (in 16 out of 25 cases), as we already expected. And, concerning the mean quantile score, the LSTM estimator performs best, closely followed by the GARCH-t estimator. 

In summary, the LSTM estimator exhibits very conservative behaviour in the stressed scenario, quickly picking up the crisis level and requiring high capital reserves. Still, after the shock period, it quickly reverts to GARCH-type dynamics that are roughly consistent with the GARCH estimator.

\begin{table}[tp!]
\phantom{.}\hspace{-0.8cm}
\begin{tabular}{lcccccccccccc}
\toprule
\multirow{2}{*}{Data} & \multicolumn{5}{c}{ $\ER$ (in \%) } &  \multicolumn{5}{c}{   $\bar S$ ($\times 10000$)  } \\
\cmidrule(lr){2-6}\cmidrule(lr){7-11}
&  emp & u & garch-n & garch-t & lstm &  emp & u & garch-n & garch-t & lstm \\
\midrule
BM1OP2 &    {\bf 0.80} &     1.40 &     2.00 &     1.40 &     2.60  &  11.14 &     12.25 &     8.33 &{\bf 7.91} &     8.63 \\ 
BM1OP3 &         2.20 &     3.00 &     1.60 &     1.40 &{\bf 1.20}  &  10.74 &     10.76 &     6.42 &     6.39 &{\bf 6.11} \\ 
BM1OP4 &         2.20 &     3.80 &     1.60 &     1.60 &{\bf 1.00}  &  8.69 &     9.94 &     5.06 &     5.09 &{\bf 4.93}   \\ 
BM2OP1 &    {\bf 1.40} &     1.80 &     2.40 &{\bf 1.40} &{\bf 1.40}  &  11.10 &     11.66 &     7.64 &     7.27 &{\bf 6.66} \\ 
BM2OP2 &         1.60 &     3.00 &     2.00 &     1.80 &{\bf 1.40}  &  8.07 &     9.00 &     5.03 &     5.04 &{\bf 4.65}   \\ 
BM2OP3 &    {\bf 0.80} &     1.60 &     1.40 &{\bf 0.80} &{\bf 0.80}  &  8.30 &     8.17 &     4.54 &{\bf 4.50} &     5.25   \\ 
BM2OP4 &    {\bf 1.40} &     2.40 &     2.40 &     2.00 &     1.80  &  8.82 &     9.21 &     5.14 &{\bf 4.73} &     5.29   \\ 
BM2OP5 &    {\bf 1.00} &     1.40 &     1.40 &     1.20 &     1.20  &  9.59 &     9.33 &     4.95 &{\bf 4.89} &     5.49   \\ 
BM3OP1 &    {\bf 1.00} &     1.60 &     1.80 &     1.60 &     1.80  &  9.95 &     10.31 &     5.59 &{\bf 5.34} &     5.52  \\ 
BM3OP2 &         1.60 &     2.80 &     2.00 &{\bf 1.20} &     2.00  &  8.67 &     10.08 &     4.70 &     4.67 &{\bf 4.46}  \\ 
BM3OP3 &    {\bf 1.20} &     2.20 &     2.20 &     1.40 &     1.40  &  10.62 &     11.22 &     4.63 &     4.53 &{\bf 4.46} \\ 
BM3OP4 &    {\bf 1.00} &     1.80 &     1.40 &{\bf 1.00} &     1.60  &  10.56 &     10.39 &{\bf 5.27} &     5.31 &     8.48 \\ 
BM3OP5 &         1.40 &     1.80 &     1.20 &{\bf 1.00} &     1.60  &  15.13 &     13.46 &{\bf 7.61} &     7.67 &     7.69 \\ 
BM4OP1 &         1.40 &     2.00 &{\bf 1.00} &     0.80 &     1.20  &  9.57 &     9.98 &     4.74 &{\bf 4.62} &     4.97   \\ 
BM4OP2 &    {\bf 1.00} &     1.80 &{\bf 1.00} &     0.60 &     0.80  &  11.01 &     11.36 &     5.20 &     5.10 &{\bf 4.73} \\ 
BM4OP3 &         1.40 &     2.20 &{\bf 1.20} &{\bf 1.20} &     1.60  &  11.90 &     12.25 &     5.50 &     5.41 &{\bf 5.30} \\ 
BM4OP4 &    {\bf 1.20} &     1.80 &     0.20 &     0.20 &     0.60  &  12.32 &     12.91 &     5.80 &     5.70 &{\bf 5.55} \\ 
BM4OP5 &         1.80 &     1.80 &     1.40 &{\bf 1.00} &     0.80  &  14.73 &     14.82 &     8.21 &     8.28 &{\bf 7.87} \\ 
BM5OP2 &    {\bf 1.20} &     1.80 &{\bf 1.20} &{\bf 1.20} &     0.60  &  14.39 &     14.62 &     7.30 &{\bf 7.17} &     7.64 \\ 
BM5OP3 &    {\bf 1.00} &     1.80 &     0.80 &     0.60 &{\bf 1.00}  &  13.83 &     14.05 &{\bf 7.65} &     7.75 &     8.02 \\ 
BM5OP4 &    {\bf 0.80} &     1.60 &     1.60 &     1.40 &     2.40  &  13.71 &     13.48 &     9.04 &{\bf 8.53} &     8.71 \\ 
HiBMHiOP &  {\bf 1.20} &     2.00 &     1.60 &     1.40 &     3.40  &    13.72 &     15.71 &     9.74 &{\bf 9.18} &  17.54 \\ 
HiBMLoOP &       1.80 &     2.20 &     1.60 &{\bf 1.20} &{\bf 1.20}  &    15.32 &     15.19 &     6.79 &{\bf 6.75} &   7.33 \\ 
LoBMHiOP &  {\bf 1.00} &     2.40 &     1.40 &     1.40 &     2.00  &    8.22 &     8.38 &     4.99 &     5.05 &{\bf 4.98} \\ 
LoBMLoOP &  {\bf 1.20} &     3.00 &     2.40 &     2.40 &     1.60  &    11.02 &     10.78 &     7.96 &   7.82 &{\bf 7.26} \\ 
\midrule
Bold count &   16 &0 &4 &9 &7  & 0 &0 &3 &10 &12  \\
\bottomrule
\end{tabular} 
\caption{\emph{Stressed period:} The  \emph{exception rate} (top) and the \emph{mean quantile score} (bottom) for estimation $\var_{1\%}$ on the Fama \& French data set with  estimation window length $n=250$ for various estimators (empirical, unbiased, GARCH-n, GARCH-t and LSTM). For the exception rate (ER), values closest to $1\%$ perform best (marked in bold). As summary statistics we provide the bold count. For the mean quantile score, lowest values perform best. }
\label{T:fama2}  
\end{table}

\section{Conclusion}\label{S:conclusions}

Our results indicate that even for a relatively short time series, the LSTM 
may serve as an efficient and highly sensitive estimator of value-at-risk. As is typical in finance, we face a small data problem with observation length of 250 or 50. In light of this, it is not particularly surprising that a shallow LSTM performs best. 
The estimators were evaluated on simulated and market data and, while on simulated data with GARCH dynamics the GARCH estimators outperform all other estimators (with LSTM being very close in many cases), this does not translate to market data. 

On real market data the LSTM is more sensitive towards increasing or decreasing volatility and outperforms all existing estimators in terms of exception rate and mean quantile score.

The results in this paper represent a promising step towards the efficient application of machine learning methods for the estimation of risk. 
While the framework proposed in this paper is tailored only to the portfolio level, it can be applied directly to risk monitoring. 
For example, given historical P\&Ls, the  base estimator could be compared to the LSTM  in order to check whether the risk is correctly estimated. Also, by including the base estimator output in the dataset for training the  LSTM, one can easily determine whether the production model can be improved. Plus, its  ability to adapt to changing market environments can be checked using this method. While this study concentrates on value-at-risk, further risk measures like expected shortfall could be explored in future work.

\section*{Acknowledgements}
The first author, Weronika Ormaniec, acknowledges partial support by Project operated within MEiN Programme,  MNiSW/2019/395/DIR/KH (“Szkola Orlow”), co-financed by the EU European Social Fund, Operational Program Knowledge Education Development. The second author, Marcin Pitera, acknowledges support from the National Science Centre, Poland, via project 2020/37/B/HS4/00120.

\end{document}